\documentclass[10pt, letter]{article}

\pdfoutput=1

\usepackage{amsmath, amsthm, amssymb, amsfonts}
\usepackage{bbm}
\usepackage{bbold}
\usepackage[basic]{complexity}
\usepackage[pdftex,bookmarks,colorlinks]{hyperref}
\usepackage{enumitem}

\newtheorem{theorem}{Theorem}
\newtheorem{lemma}[theorem]{Lemma}
\newtheorem{example}{Example}
\newtheorem{defn}{Definition}

\newcommand{\sectionref}[1]{Section \ref{#1}}

\DeclareMathOperator*{\argmin}{arg\,min}

\newcommand{\boldVar}[1]{\mathbf{#1}} 
\newcommand{\mvar}[1]{\boldVar{#1}} 
\newcommand{\vvar}[1]{\vec{#1}} 

\newcommand{\innerprod}[2]{\langle #1, #2 \rangle}
\newcommand{\innerprodOma}[2]{\innerprod{#1}{#2}_{1 - \alpha}}
\newcommand{\innerprodA}[2]{\innerprod{#1}{#2}_{\A}}
\renewcommand{\E}{\mathbb{E}} 
\newcommand{\Eof}[1]{\E\left[#1\right]}
\newcommand{\EikOf}[1]{\E_{i_k}\left[#1\right]}

\newcommand{\A}{\mvar{A}}
\newcommand{\ma}{\A}
\newcommand{\mb}{\mvar{B}}
\newcommand{\Pmat}{\mvar{P}}
\newcommand{\fun}{f}
\newcommand{\f}{\fun}
\newcommand{\x}{\vvar{x}}
\newcommand{\vx}{\vvar{x}}

\newcommand{\y}{\vvar{y}}
\newcommand{\vy}{\vvar{y}}
\newcommand{\vytilde}{\tilde{y}}
\newcommand{\vz}{\vvar{z}}
\newcommand{\vb}{\vvar{b}}
\newcommand{\vc}{\vvar{c}}
\newcommand{\vv}{\vvar{v}}
\newcommand{\vs}{\vvar{s}}
\newcommand{\vzero}{\vvar{0}}
\newcommand{\xinit}{\x_0}
\newcommand{\xopt}{\x^*}
\newcommand{\optValue}{\fun^*}
\newcommand{\optF}{\optValue}
\newcommand{\optPoint}{\x^*}
\newcommand{\optX}{\optPoint}

\newcommand{\gradfiVal}{\f_i}
\newcommand{\gradfiVec}{\vec{\f}_i}
\newcommand{\gradfikVal}{\f_{i_k}}
\newcommand{\gradfikVec}{\vec{\f}_{i_k}}
\newcommand{\bvec}{\vvar{b}}

\newcommand{\infSeq}[1]{\{#1\}_{k = 0}^{\infty}}
\newcommand{\seqInit}[1]{{#1}_{0}}
\newcommand{\seqCurr}[1]{{#1}_{k}}
\newcommand{\seqNext}[1]{{#1}_{k + 1}}

\newcommand{\vvInit}{\seqInit{\vv}}
\newcommand{\vvCurr}{\seqCurr{\vv}}
\newcommand{\vvNext}{\seqNext{\vv}}

\newcommand{\vxInit}{\seqInit{\vx}}
\newcommand{\vxCurr}{\seqCurr{\vx}}
\newcommand{\vxNext}{\seqNext{\vx}}

\newcommand{\vyInit}{\seqInit{\vy}}
\newcommand{\vyCurr}{\seqCurr{\vy}}
\newcommand{\vyNext}{\seqNext{\vy}}

\newcommand{\phiInit}{\seqInit{\phi}}
\newcommand{\phiCurr}{\seqCurr{\phi}}
\newcommand{\phiNext}{\seqNext{\phi}}

\newcommand{\etaInit}{\seqInit{\eta}}
\newcommand{\etaCurr}{\seqCurr{\eta}}
\newcommand{\etaNext}{\seqNext{\eta}}

\newcommand{\zetaInit}{\seqInit{\zeta}}
\newcommand{\zetaCurr}{\seqCurr{\zeta}}
\newcommand{\zetaNext}{\seqNext{\zeta}}

\newcommand{\thetaCurr}{\seqCurr{\theta}}

\newcommand{\indexCurr}{\seqCurr{i}}

\newcommand{\probCurr}{p_{\indexCurr}}

\newcommand{\Ecurr}{\seqCurr{\E}}
\newcommand{\Enext}{\seqNext{\E}}
\newcommand{\EcurrOf}[1]{\Ecurr\left[#1\right]}
\newcommand{\EnextOf}[1]{\Enext\left[#1\right]}

\newcommand{\verrOneCurr}{\vvar{\epsilon}_{1,k}}
\newcommand{\verrTwoCurr}{\vvar{\epsilon}_{2,k}}

\newcommand{\nnz}{\mathrm{nnz}}
\newcommand{\proj}{\mathrm{proj}}

\newcommand{\constL}{L}
\newcommand{\constLi}{L_i}
\newcommand{\constConv}{\mu}

\newcommand{\basisI}{\vvar{e}_i}
\newcommand{\onesVec}{\mathbb{1}}

\newcommand{\indicVec}[1]{\onesVec_{#1}}

\newcommand{\defeq}{\stackrel{\mathrm{\scriptscriptstyle def}}{=}}

\newcommand{\demands}{\vvar{\chi}}

\newcommand{\lap}{\mvar{\mathcal L}}
\newcommand{\pseudo}[1]{{#1}^\dagger}
\newcommand{\lapPseudo}{\pseudo{\lap}}
\newcommand{\incMatrix}{\mvar{B}}
\newcommand{\rMatrix}{\mvar{R}}

\newcommand{\cycleMatrix}{\mvar{C}}

\newcommand{\tree}{T}

\newcommand{\treeCycle}[1]{C_{#1}}
\newcommand{\treeCycleVec}[1]{\vvar{c}_{#1}}
\newcommand{\st}{\mathrm{st}}
\newcommand{\stretchEdge}[1]{\mathrm{st}\left(#1\right)}

\newcommand{\offtreeEdgeSet}{E \setminus \tree}

\newcommand{\gradient}{\nabla}
\newcommand{\grad}{\gradient}
\newcommand{\hessian}{\grad^2}
\newcommand{\norm}[1]{\left\|#1\right\|}
\newcommand{\normTwo}[1]{\norm{#1}_{2}}
\newcommand{\normFro}[1]{\norm{#1}_F}
\newcommand{\normR}[1]{\norm{#1}_{\rMatrix}}
\newcommand{\normL}[1]{\norm{#1}_{\lap}}
\newcommand{\normOma}[1]{\norm{#1}_{1 - \alpha}}
\newcommand{\normOmaDual}[1]{\norm{#1}_{1 - \alpha}^*}
\newcommand{\normA}[1]{\norm{#1}_{\A}}

\newcommand{\sigmaOma}{\sigma_{1 - \alpha}}
\newcommand{\sigmaA}{\sigma_{\A}}

\renewcommand{\R}{\mathbb{R}} 
\newcommand{\Rn}{\R^{n}}
\newcommand{\Rm}{\R^{m}}
\newcommand{\Rnn}{\R^{n \times n}}

\newcommand{\Redgevec}{\R^{E}}

\newcommand{\Rev}{\R^{E \times V}}

\newcommand{\Roffedgevec}{\R^{\offtreeEdgeSet}}
\newcommand{\RnToR}{\R^{n} \rightarrow \R}

\newcommand{\tLi}{\tilde{L}_{i}}
\newcommand{\tLik}{\tilde{L}_{i_{k}}}
\newcommand{\tlik}{\tilde{L}_{i_{k}}}
\newcommand{\tSa}{\tilde{S}_{\alpha}}
\newcommand{\tsa}{\tilde{S}_{\alpha}}

\newcommand{\tr}{\mathrm{tr}}

\newcommand{\sddRuntimeOurs}{O(m \log^{3/2} n \sqrt{\log \log n} \log (\frac{\log n}{\epsilon}))}
\newcommand{\sddRuntimeOursFlow}{O(m \log^{3/2} n \sqrt{\log \log n} \log (\frac{1}{\epsilon}))}

\begin{document}

\title{Efficient Accelerated Coordinate Descent Methods\\
and \\
Faster Algorithms for Solving Linear Systems}

\author{
    Yin Tat Lee\\
    MIT\\
    yintat@mit.edu
  \and
    Aaron Sidford\\
    MIT\\
    sidford@mit.edu
}

\date{}

\maketitle

\begin{abstract}
In this paper we show how to accelerate randomized coordinate descent methods and achieve faster convergence rates without paying per-iteration costs in asymptotic running time. In particular, we show how to generalize and efficiently implement a method proposed by Nesterov, giving faster asymptotic running times for various algorithms that use standard coordinate descent as a black box. In addition to providing a proof of convergence for this new general method, we show that it is numerically stable, efficiently implementable, and in certain regimes, asymptotically optimal. 

To highlight the computational power of this algorithm, we show how it can used to create faster linear system solvers in several regimes:
\begin{itemize}

  \item We show how this method achieves a faster asymptotic runtime than conjugate gradient for solving a broad class of symmetric positive definite systems of equations. 
  
  \item We improve the best known asymptotic convergence guarantees for Kaczmarz methods, a popular technique for image reconstruction and solving overdetermined systems of equations, by accelerating a randomized algorithm of Strohmer and Vershynin.
  
  \item We achieve the best known running time for solving Symmetric Diagonally Dominant (SDD) system of equations in the unit-cost RAM model, obtaining an $\sddRuntimeOurs$ asymptotic running time by accelerating a recent solver by Kelner \emph{et al.}
\end{itemize}

Beyond the independent interest of these solvers, we believe they highlight the versatility of the approach of this paper and we hope that they will open the door for further algorithmic improvements in the future.
\end{abstract}
\newpage

\section{Introduction}
In recent years iterative methods for convex optimization that make progress in sublinear time using only partial information about the function and its gradient have become of increased importance to both the theory and practice of computer science. From a practical perspective, the increasing volume and distributed nature of data are forcing efficient practical algorithms to be amenable to asynchronous and parallel settings where only a subset of the data is available to a single processor at any point in time. From a theoretical perspective, rapidly converging algorithms with sublinear time update steps create hope for new provable asymptotic running times for old problems and stronger guarantees for efficient algorithms in distributed and asynchronous settings. 

The idea of using simple sublinear-time iterative steps to solve convex optimization problems is an old one \cite{Kaczmarz, zangwill1969, Bauschke:1996:PAS:240441.240442}. It is an algorithmic design principle that has seen great practical success \cite{Natterer, herman2009fundamentals, Bauschke:1996:PAS:240441.240442} but has been notoriously difficult to analyze. In the past few years great strides have been taken towards developing a theoretical understanding of randomized variants of these approaches. Of particular relevance to this paper, in 2006 Strohmer and Vershynin \cite{Strohmer2008} showed that a particular sublinear update algorithm for solving overconstrained linear systems called \emph{randomized Kaczmarz} converges exponentially, in 2010 Nesterov \cite{Nesterov2012} analyzed randomized analog of gradient descent
that updates only a single coordinate in each iteration, called \emph{coordinate gradient descent method}, and provided a computationally inefficient but theoretically interesting accelerated variant, called \emph{accelerated coordinate gradient descent method (ACDM)}, and in 2013 Kelner et al. \cite{Kelner2013} presented a simple combinatorial iterative algorithm with sublinear-time update steps that can be used to solve \emph{symmetric diagonally dominant (SDD)} linear systems, a broad class of linear systems with numerous applications.

In this paper we provide a framework that both strengthens and unifies these results. We present a more general version of Nesterov's ACDM and show how to implement it so that each iteration has the same asymptotic runtime as its non-accelerated variants. We show that this method is numerically stable and optimal under certain assumptions. Then we show how to use this method to outperform conjugate gradient in solving a general class of symmetric positive definite systems of equations. Furthermore, we show how to cast both randomized Kaczmarz and the SDD solver of Kelner et al. in this framework and achieve faster running times through the use of ACDM. 

Due to the success of the Kaczmarz method in practice \cite{Natterer, herman2009fundamentals} and due to the wide array of theoretical problems for which the fastest running time is obtained through the use of a nearly linear time SDD solver \cite{KelnerMadry, christiano2011electrical, Spielman:2012:AGT:2359888.2359901}, we hope that the ideas in this paper can be used to make advancements on both fronts.

\subsection{Previous Work}

Gradient descent is one of the oldest and most fundamental methods in convex optimization. Given a convex differentiable function the \emph{gradient descent method} is a simple greedy iterative method that computes the gradient at the current point and uses that information to perform an update and make progress. This method is central to much of scientific computing and from a theoretical perspective the standard method is well understood \cite{Nesterov2003}. There are multiple more sophisticated variants of this method \cite{kelley1987iterative}, but many of them have only estimates of local convergence rates which makes them difficult to be applied to theoretical problems and be compared in general.

In 1983, Nesterov \cite{Nesterov1983} proposed a way to accelerate the gradient descent method by iteratively developing an approximation to the function through what he calls an \emph{estimate sequence}. This \emph{accelerated gradient descent method} or \emph{fast gradient method} has the same worst case running time as conjugate gradient method and it is applicable to general convex functions. Recently, this method has been used to improve the fastest known running time of some fundamental problems in computer science, such as compressive sensing \cite{Beck2009,Becker2011}, undirected maximum flow \cite{christiano2011electrical,Lee2013,kelner2013almost}, linear programming \cite{Nesterov:2008:RCS:1451525.1451532,Bienstock:2004:SFP:1007352.1007382}.

The accelerated gradient descent method is known to achieve an optimal (up to constants) convergence rate among all first order methods, that is algorithm that only have access to the function's value and gradient \cite{Nesterov2003}.  Therefore, to further improve accelerated gradient descent one must either assume more information about the function or find a way to reduce the cost of each iteration. Using the idea of fast but crude iteration steps, Nesterov proposed a randomized coordinate descent method \cite{Nesterov2012}, which minimizes convex functions by updating one randomly chosen coordinate in each iteration.

Coordinate descent methods, which use gradient information about a single coordinate to update a single coordinate in each iteration, have been around for a long time \cite{zangwill1969}. Various schemes have been considered for picking the coordinate to update, such as cyclic coordinate update and the best coordinate update, however these schemes are either hard to estimate \cite{Luo1992} or difficult to be implemented efficiently. 
Both the recent work of Strohmer and Vershynin \cite{Strohmer2008} and Nesterov \cite{Nesterov2012} overcame these obstacles by showing that by performing particular randomized updates one can produce methods with provable global convergence rate and small costs per iteration. 

Applying the similar ideas of accelerated gradient descent, Nesterov also proposed an accelerated variant called the \emph{accelerated coordinate descent method (ACDM)} that achieves a faster convergence rate while still only considering a single coordinate of the gradient at a time. However, in both Nesterov's paper \cite{Nesterov2012} and later work \cite{Richtarik2011}, this method was considered inefficient as the computational complexity of the naive implementation of each iteration of ACDM requires $\Theta(n)$ time to update every coordinate of the input, at which point the accelerated gradient descent method would seem preferable.

\subsection{Our Contributions}

In this paper, we generalize Nesterov's ACDM and present a simple technique to implement each coordinate update step efficiently. Our contributions towards understanding ACDM include the following:
\begin{itemize}
  \item \textbf{Generalization:} A generalization of ACDM to a broader class of sampling probabilities, overcoming technical challenges due to skewed sampling probabilities, so that any convergence rate achieved through Nesterov's coordinate descent method can be improved by ACDM. This generalization was essential for applications considered later in the paper.
  \item \textbf{Probabilistic Estimate Sequences:} A self contained proof of correctness for ACDM motivated by a generalization of estimate sequences we simply call \emph{probabilistic estimate sequences}.
  \item \textbf{Efficiency:} A proof that under mild assumptions about the oracle for querying function and gradient values, each iteration of ACDM can be implemented with the same asymptotic cost as an equivalent coordinate descent step.  
  \item \textbf{Numerical Stability:} A proof that ACDM is numerically stable and can be implemented with finite precision arithmetic and no overhead in the standard unit cost RAM model.
  \item \textbf{Lower Bound:} A lower bound argument showing that ACDM achieves an optimal convergence rate among a certain class of coordinate descent algorithms.
\end{itemize}

In some sense, the principle difference between the asymptotic running time of ACDM and accelerated gradient descent (or conjugate gradient in the linear system case) is that as accelerated gradient descent depends on the maximum eigenvalue of the Hessian of the function being minimized, ACDM instead depends on the trace of the Hessian and has the possibility of each iteration costing a small fraction of the cost a single iteration of of accelerated gradient descent. As a result, any nontrivial bound on the trace of the Hessian and the computational complexity of performing a single coordinate update creates the opportunity for ACDM to yield improved running times. To emphasize this point, we focus on applications of our method to solving linear systems and we use three different cases to illustrate the flexibility and competitiveness of our method. 

\begin{itemize}
  \item \textbf{Symmetric Positive Definite Systems:} We show that under mild assumptions ACDM solves positive definite systems with a faster asymptotic running time than conjugate gradient (and an even milder set of assumptions for Chebyshev method), and it is an asymptotically optimal algorithm for solving general systems in certain regimes. 
  \item \textbf{Overdetermined Systems:} For over-constrained systems of equations the randomized Kaczmarz method of Strohmer and Vershynin \cite{Strohmer2008}, which iteratively picks a random constraint and projects the current solution onto the plane corresponding to a random constraint, has been shown to have strong convergence guarantees and appealing practical performance. We show how to cast this method in the framework of coordinate descent and accelerate it using ACDM yielding improved asymptotic performance. Given the appeal of Kaczmarz methods for practical applications such as image reconstructions \cite{herman2009fundamentals} , there is hope that this could yield improved performance in practice. 

    \item \textbf{Symmetric Diagonally Dominant (SDD) Systems:} Symmetric diagonally dominant systems of equations are a common class of systems with broad applications ranging from finite element simulations \cite{boman2008solving} to computer vision \cite{DBLP:journals/cviu/KoutisMT11} and beyond. Furthermore, since a breakthrough result in 2004 by Spielman and Teng \cite{spielman2004nearly} showed that such systems can be solved in nearly-linear time, such systems have been used to create the fastest algorithm for a variety of problems ranging from max flow \cite{kelner2013almost} to sampling random spanning trees \cite{KelnerMadry} and much more. The fastest known solver for these systems in the standard unit-cost RAM model is due to Kelner \emph{et. al.} \cite{Kelner2013} is  $\tilde{O}(m\log^{2}n \log \frac{1}{\epsilon})$ where $m$ is the number of nonzero entries in the matrix, the matrix is $n \times n$, and we use $\tilde{O}$ to hide $O(\poly(\log \log n))$ terms. We show how direct application of ACDM to a simple algorithm in \cite{Kelner2013} yields a faster SDD solver with an asymptotic runtime of $\tilde{O}(m \log^{1.5}n \log \frac{1}{\epsilon})$ in the unit-cost RAM model, getting closer to the fastest known running time of $\tilde{O}(m \log n \log \frac{1}{\epsilon})$ by Koutis, Miller, and Peng \cite{KMP11} in a less restrictive computational model.
\end{itemize}

We remark that while our applications analysis focus on solving linear systems there is hope that our ACDM algorithm will have a broader impact in both theory and practice of efficient algorithms. Just as the accelerated gradient descent method has improved the theoretical and empirical running time of various, both linear and nonlinear, gradient descent algorithms \cite{goldstein2012fast, Beck2009}, we hope that ACDM will improve the running time of various algorithms for which coordinate descent based approaches have proven effective. Given the generality of our analysis and the previous difficulty in analyzing such methods, we hope that this is just the next towards a new class of provably efficient algorithms with good empirical performance.

\subsection{Paper Organization}

The rest of our paper is organized as follows. In \sectionref{sec:preliminaries}, we introduce the problem and function properties that we will use for optimization. In \sectionref{sec:iter_methods}, we briefly review the mathematics behind gradient descent, accelerated gradient descent, and coordinate descent. In \sectionref{sec:acdm_upper}, we present our general ACDM implementation, prove correctness, and show how to implement the update steps efficiently. In \sectionref{sec:applications}, we show how to apply ACDM to achieve faster runtimes for various linear system solving scenarios. In \sectionref{sec:acdm_lower}, we give lower bounds that show that ACDM is optimal in certain regimes, and in the appendix we provide missing details of the ACDM convergence and numerical stability proofs.
\section{Preliminaries}
\label{sec:preliminaries}

In this paper, we consider the unconstrained minimization problem \footnote{Many of the results in this paper can be generalized to constrained minimization problems \cite{Nesterov2012}, problems where $\fun$ is strongly convex with respect to different norms \cite{Nesterov2012}, and problems where each coordinate is a higher dimension variable (i.e. the block setting) \cite{Richtarik2011}. However, for simplicity we focus on the basic coordinate unconstrained problem in the Euclidian norm.
}
\[
\min_{\x \in \Rn} \fun(\x)
\]
where the \emph{objective function} $\fun : \Rn \rightarrow \R$ is continuously differentiable and convex, meaning that
\[
\forall \x, \y \in \Rn 
\enspace : \enspace
\f(\y) \geq \f(\x) + \innerprod{\gradient \f(x)}{\y - \x}
\enspace.
\]
We let $\f^{*} \defeq \min_{\x \in \Rn} \fun(\x)$ denote the minimum value of this optimization problem and we let $\x^{*} \defeq \argmin_{\x \in \Rn} \fun(\x)$ denote an arbitrary point that achieves this value.

To minimize $\fun$, we restrict our attention to \emph{first-order} iterative methods, that is algorithms that generate a sequence of points $\vx_1, \ldots, \vx_k$ such that $\lim_{k \rightarrow \infty} \f(\vx_k) = \f^*$, while only evaluating the objective function, $\fun$, and its gradient $\gradient f$, at points. In other words, other than the global function parameters related to $\f$ that we define in this section, we assume that the algorithms under consideration have no further knowledge regarding the structure of $\f$. To compare such algorithms, we say that \emph{an iterative method has convergence rate $r$} if $f(\x_{k})-f^{*}\leq O((1-r)^{k})$ for this method.

Now, we say that \emph{$\fun$ has convexity parameter $\sigma$ with respect to some norm $\norm{\cdot}$} if the following holds
\begin{equation}
\forall \x, \y \in \Rn
\enspace : \enspace
\f(\y)
\geq 
\f(\x)
+ \innerprod{\gradient \f(\x)}{\y - \x}
+\frac{\sigma}{2} \norm{\y - \x}^{2}
\label{eq:lower_env}
\end{equation}
and we say $\fun$ \emph{strongly convex} if $\sigma > 0$. We refer to the right hand side of (\ref{eq:lower_env}) as the \emph{lower envelope of $\fun$ at $\x$} and for notational convenience when the norm $\norm{\cdot}$ is not specified explicitly we assume it to be the standard Euclidian norm $\norm{x} \defeq \sqrt{\sum_{i} \x_i^2}$. 

Furthermore, we say $\f$ has $\constL$-Lipschitz gradient if 
\[
\forall \x, \y \in \Rn
\enspace : \enspace
\norm{\gradient \f(\y) - \gradient \f(\x)} \leq L \norm{\y - \x}
\]
The definition is related to an upper bound on $\f$ as follows:
\begin{lemma}
\label{lem:eqv_def_of_L}\cite[Thm 2.1.5]{Nesterov2003} 
For continuously differentiable $\f : \Rn \rightarrow \R$ and $L > 0$, it has $\constL$-Lipschitz gradient if and only if 
\begin{equation}
\label{eq:upper_env}
\forall \x, \y \in \Rn
\enspace : \enspace
f(\y) \leq f(\x) +
\left\langle \gradient f(\x), \y - \x\right\rangle 
+ \frac{L}{2}\norm{\x - \y}^{2}
\enspace.
\end{equation}
\end{lemma}
We call the right hand side of (\ref{eq:upper_env}) the \emph{upper envelope of $\f$ at $\x$}. 

The convexity parameter $\constConv$ and the Lipschitz constant of the gradient $\constL$ provide lower and upper bounds on $\f$. They serve as the essential characterization of $\f$ for first-order methods and they are typically the only information about $\f$ provided to both gradient and accelerated gradient descent methods (besides the oracle access to $\f$ and $\gradient \f$). For twice differentiable $\f$, these values can also be computed by properties of the Hessian of $\f$ by the following well known lemma:

\begin{lemma}[\cite{Nesterov2012}]
\label{lem:twicediff_prop}
Twice differentiable $\f : \Rn \rightarrow \R$ has convexity parameter $\mu$ and $\constL$-Lipschitz gradient with respect to norm $\norm{\cdot}$ if and only if $\forall \x \in \Rn$ the Hessian of $\f$ at $\x$, $\gradient^2 \f(\x) \in \Rnn$ satisfies 
\[
\forall \vy \in \Rn
\enspace : \enspace
\mu \norm{\vy}^2
\leq
\vy^T \left(\gradient^2 \f(\x)\right) \vy
\leq
L \norm{\vy}^2
\]
\end{lemma}

To analyze \emph{coordinate-based iterative methods}, that is iterative methods that only consider one component of the current point or current gradient in each iteration, we need to define several additional parameters characterizing $\f$. For all $i \in [n]$, let $\basisI \in \Rn$ denote the standard basis vector for coordinate $i$, let $\gradfiVal(\x) \in \Rn$ denote the partial derivative of $\f$ at $\x$ along $\basisI$, i.e. $\gradfiVal(\x) \defeq \basisI^T \gradient\f(\x)$, and let $\gradfiVec(\x)$ denote the corresponding vector, i.e $\gradfiVec(\x) \defeq \gradfiVal \cdot \basisI$. We say that \emph{$\f$ has component-wise Lipschitz continuous gradient with Lipschitz constants $\{L_{i}\}$} if 
\[
\forall \x \in \Rn, \enspace
\forall t \in \R, \enspace
\forall i \in [n]
\enspace : \enspace
|\gradfiVal(\x + t \cdot \basisI)- \gradfiVal(\x)| \leq \constLi \cdot |t|
\]
and for all $\alpha \geq 0$ we let $S_\alpha \defeq \sum_{i=1}^{n} L_{i}^{\alpha}$ denote the total component-wise Lipschitz constant of $\gradient f$. Later we will see that $S_\alpha$ has a similar role for coordinate descent as $\constL$ has for gradient descent.

We give two examples for convex functions induced by linear systems and calculate their parameters. Note that even though one example can be deduced from the other, we provide both as the analysis allows us to introduce more notation.

\begin{example}
\label{psd_example}
Let $\f(\x) \defeq \frac{1}{2} \innerprod{\A \x}{\x} - \innerprod{\x}{\bvec}$ for symmetric positive definite matrix $\A \in \R^{n \times n}$.  Since $\A = \A^T$ clearly $\grad \f(\x) = \A \x - \bvec$ and $\grad^2 \f(\x) = \A$. Therefore, by 
Lemma \ref{lem:twicediff_prop}, $L$ and $\sigma$ satisfy
\[
\sigma \norm{\x}^2 \leq \x^T \A \x \leq L \norm{\x}^2
\enspace.
\]
Consequently, $\sigma$ is the the smallest eigenvalue $\lambda_{\min}$ of $\A$ and $L$ is the largest eigenvalue $\lambda_{\max}$ of $\A$. Furthermore, $\forall i \in [n]$ we see that $f_{i}(\x)= \vvar{e}_{i}^{T}\left(\A \vx - \vb\right)$ and therefore $L_i$ satisfies
\[
\forall t \in \R
\enspace : \enspace
|t| \cdot |\A_{ii}|=|\vvar{e}_{i}^{T}\A(te_{i})|\leq L_{i}|t|
\enspace.
\]
Since the positive definiteness of $\A$ implies that $\A$ is positive on diagonal, we have $L_{i} = \A_{ii}$, and consequently
$
S_{1} = \tr(\A) = \sum_{i = 1}^{n} \A_{ii} = \sum_{i=1}^{n}\lambda_{i}
$
where $\lambda_{i}$ are eigenvalues of $\A$.
\end{example}

\begin{example}
\label{linear_system_example}Let $\f(\x)=\frac{1}{2}\norm{\A \x - b}^2$
for any matrix $\A$. Then $\gradient \f(\x) = \A^{T}\left(\A \vx - \vb\right)$ and $\gradient^2 \f(\x) = \A^T \A$. Hence, $\sigma$ and $L$ satisfy
\[
\sigma \norm{\x}^2 \leq \x^T \A^T \A \x \leq L \norm{\x}^2
\]
and we see that $\sigma$ is the the smallest eigenvalue $\lambda_{\min}$
of $\A^{T} \A$ and $L$ is the largest eigenvalue $\lambda_{\max}$
of $\A^{T}\A$. As in the previous example, we therefore have
$
L_{i} = \norm{a_{i}}^{2}
$
where $a_{i}$ is the $i$-th column of $\A$ and 
$S_{1}= \sum \norm{a_{i}}^{2} = \normFro{\A}^{2}$, the Frobenius norm of $\A$.
\end{example}

\section{Review of Previous Iterative Methods}
\label{sec:iter_methods}

In this section, we briefly review several standard iterative first-order method for smooth convex minimization. This overview is by no means all-inclusive, our goal is simply to familiarize the reader with numerical techniques we will make heavy use of later and motivate our presentation of the accelerated coordinate descent method. For a more comprehensive review, there are multiple good references, e.g. \cite{Nesterov2003}, \cite{Boyd:2004:CO:993483}. 

In \sectionref{sec:gradient_descent}, we briefly review the standard gradient descent method. In \sectionref{sec:accelerated_gradient_descent}, we show how to improve gradient descent by motivating and reviewing Nesterov's \emph{accelerated gradient descent method} \cite{Nesterov1983} through a more recent presentation of his via \emph{estimate sequences} \cite{Nesterov2003}. In \sectionref{sec:coordinate_descent}, we review Nesterov's \emph{coordinate gradient descent method} \cite{Nesterov2012}. In the next section we combine these concepts to present a general and efficient accelerated coordinate descent scheme.

\subsection{Gradient Descent}
\label{sec:gradient_descent}

Given an initial point $\xinit \in \Rn$ and step sizes $h_k \in \R$, the \emph{gradient descent method} applies the following simple iterative update rule:
\[
\forall k \geq 0
\enspace : \enspace
\x_{k+1} := \x_{k} - h_{k} \gradient \f(\x_{k}).
\]
For $h_{k} = \frac{1}{L}$, this method simply chooses the minimum point of the upper envelope of $\f$ at $\x_{k}$:
\[
\x_{k+1}
=
\argmin_{\y}
\left\{ 
f(\x_{k})
+
\left\langle \gradient f(\x_{k}),\y-\x_{k}\right\rangle +\frac{L}{2}\norm{\y-\x_{k}}^{2}\right\} .
\]
Thus, we see that the gradient descent method is a greedy method that chooses the minimum point based on the worst case estimate of the function based on the value of $\fun(x_k)$ and $\gradient \fun(x_{k})$. It is well known that it provides the following guarantee \cite[Cor 2.1.2, Thm 2.1.15]{Nesterov2003} 
\begin{equation}
f(\x_{k}) - \optValue 
\leq 
\frac{L}{2} \cdot 
\min\left\{ \left(1-\frac{\sigma}{L}\right)^{k},\frac{4}{k+4}\right\} 
\norm{\x_0 - \optPoint}^{2}
\enspace.
\end{equation}

\subsection{Accelerated Gradient Descent}
\label{sec:accelerated_gradient_descent}

To speed up the greedy and memory-less gradient descent method, one could make better use of the history and pick the next step to be the smallest point in  upper envelope of all points computed. Formally, one could try the following update rule
\[
\vx_{k+1} :=
\argmin_{\vy \in \Rn} 
\min_{\sum_{k=1}^{n} t_k \vec{z}_k = \vy}
\left\{ 
\sum_{k = 1}^{n} t_k \left(
f(\vx_{k}) +
\innerprod{\gradient \f(\vx_{k})}{\vec{z}_k- \vx_{k}} +
\frac{L}{2} \norm{ \vec{z}_k - \vx_{k}}^{2}
\right)
\right\}. 
\]
However, this problem is difficult to solve efficiently and requires storing all previous points. To overcome this problem, Nesterov \cite{Nesterov1983,Nesterov2003} suggested to use a quadratic function to estimate the function. Formally, we define an \emph{estimate sequence} as follows: \footnote{Note that our definition deviates slightly from Nesterov's \cite[Def 2.2.1]{Nesterov2003} in that we include condition \ref{eq:upper_estimate_sequence}.}

\begin{defn}[Estimate Sequence]
A triple of sequences $\{\phi_{k}(x),\eta_{k},\vx_{k}\}_{k=0}^{\infty}$
is called an \emph{estimate sequence} of $\f$ if $\lim_{k \rightarrow \infty} \eta_{k} = 0$ and for any $\x \in \Rn$ and $k \geq 0$ we have 
\begin{equation}
\phi_{k}(\vx)\leq(1-\eta_{k})f(\vx)+\eta_{k}\phi_{0}(\vx)\label{eq:lower_estimate_sequence}
\end{equation}
and
\begin{equation}
f(\vx_{k})\leq\min_{\vx\in\Rn}\phi_{k}(\vx).\label{eq:upper_estimate_sequence}
\end{equation}
\end{defn}

An estimate sequence of $\fun$ is an approximate lower bound of
$\fun$ which is slightly above $\fun^{*}$. This relaxed definition allows
us to find a better computable approximation of $\f$ instead of relying
on the worst case upper envelope at each step. 

A good estimate sequence gives an efficient algorithm \cite[Lem 2.2.1]{Nesterov2003} by the following
\[
\lim_{k \rightarrow \infty} \fun(\x_{k}) - \optValue
\leq
\lim_{k \rightarrow \infty} \eta_{k} \left(\phi_{0}(\x^{*})-\f^{*}\right)
= 0.
\]
Since an estimate sequence is an approximate lower bound, a natural
computable candidate is to use the convex combination of lower envelopes
of $\fun$ at some points. 

Since it can be shown that any convex combinations of lower envelopes at evaluation points $\{y_{k}\}$ satisfies (\ref{eq:lower_estimate_sequence}) under some mild condition, additional points $\{y_{k}\}$ other than $\{x_{k}\}$ can be used to tune the algorithm. Nesterov's \emph{accelerated gradient descent method} can be obtained by tuning the the free parameters $\{y_{k}\}$ and $\{\eta_{k}\}$ to satisfy \eqref{eq:upper_estimate_sequence}. Among all first order methods, this method is optimal up to constants in terms of number of queries made to $\fun$ and $\gradient f$. The performance of the accelerated gradient descent method can be characterized as follows: \cite{Nesterov2003}
\begin{equation}
\label{eq:AGDM}
f(\x_{k}) - \optValue 
\leq 
L \cdot 
\min\left\{ \left(1-\sqrt{\frac{\sigma}{L}}\right)^{k},\frac{4}{(k+2)^{2}}\right\} 
\norm{\x_0 - \optPoint}^{2}
\enspace.
\end{equation}

\subsection{Coordinate Descent}
\label{sec:coordinate_descent}

The \emph{coordinate descent method} of Nesterov \cite{Nesterov2012} is a variant of gradient descent in which only one coordinate of the current iterate is updated at a time. For a fixed $\alpha \in \R$, each iteration $k$ of coordinate descent consists of picking a random a random coordinate $i_k \in [n]$ where
\[
\Pr[i_k = j] = P_\alpha(j) 
\enspace \text{ where } \enspace 
P_\alpha(j) \defeq \frac{L_{i_k}^\alpha}{S_\alpha}
\]
and then performing a gradient descent step on that coordinate:
\[
\vx_{k + 1} := \vx_k - \frac{1}{L_{i_k}} \vvar{f}_{i_k}(\vx_{k}).
\]

To analyze this algorithm's convergence rate, we define the norm $\normOma{\cdot}$, its dual $\normOma{\cdot}^*$, and the inner product $\innerprod{\cdot}{\cdot}_{1 - \alpha}$ which induces this norm as follows:
\[
\normOma{\vx} \defeq \sqrt{\sum_{i = 1}^{n} L_i^{1 - \alpha} \vx_i^2}
\enspace \text{ and }
\normOma{\vx}^* \defeq \sqrt{\sum_{i = 1}^{n} L_i^{-(1 - \alpha)} \vx_i^2}
\enspace \text{ and }
\innerprod{\vx}{\vy}_{1 - \alpha} \defeq \sum_{i = 1}^{n} L_i^{1- \alpha} \vx_i \vy_i
\]
and we let $\sigma_{1 - \alpha}$ denote the convexity parameter of $f$ with respect to $\normOma{\cdot}$.

Using the definition of coordinate-wise Lipschitz constant, each step can be shown to have the following guarantee on expected improvement \cite{Nesterov2012}
\[
f(\x_{k}) - \E\left[\f(\x_{k+1})\right]
\geq
\frac{1}{2S_{\alpha}}\left(\normOmaDual{\grad \f(\x_k)}\right)^{2}.
\]
and further analysis shows the following convergence guarantee coordinate descent \cite{Nesterov2012}
\[
\E\left[\f(\vx_k)\right] - \optValue
\leq \min 
\left\{
\frac{2 S_\alpha}{k + 4} \left(
\max_{f(\vy) \leq f(\vx_0)}
 \normOma{\vy - \optPoint}^2\right)
\enspace , \enspace
\left(1 - \frac{\sigma_{1 - \alpha}}{S_\alpha}\right)^k
\left(f(\vx_0) - \optValue\right) 
\right\}.
\]

\section{General Accelerated Coordinate Descent}
\label{sec:acdm_upper}

In this section, we present our general and iteration-efficient \emph{accelerated coordinate descent method (ACDM)}. In particular, we show how to improve the asymptotic convergence rate of any coordinate descent based algorithm without paying asymptotic cost in the running time due to increased cost in querying and updating a coordinate. We remark that the bulk of the credit for conceiving of such a method belongs to Nesterov \cite{Nesterov2012} who provided a different proof of convergence for such a method for the $\alpha = 0$ case, however we note that changes to the algorithm were necessary to deal with the $\alpha = 1$ case used in all of our applications.
 
The rest of this section is structured as follows. In \sectionref{sec:prob_est_seq}, we introduce and prove the correctness of general ACDM through what we call \emph{(probabilistic) estimation sequences}, in \sectionref{sec:acdm_upper_num_stab}, we present the numerical stability results we achieve for this method, and in \sectionref{sec:efficient_iteration}, we show how to implement this method efficiently. In Appendix \ref{sec:app:acdm_coefficients}, we include some additional details for the correctness proof and in Appendix \ref{sec:app:num_stab}, we provide the details of the numerical stability proof.
 
\subsection{ACDM by Probabilistic Estimate Sequences}
\label{sec:prob_est_seq}

Following the spirit of the estimate sequence proof of accelerated gradient descent \cite{Nesterov2003}, here we present a proof of ACDM convergence through what we call a \emph{(probabilistic) estimation sequence}.

\begin{defn}[(Probabilistic) Estimate Sequence]
A triple of sequences $\infSeq{\phiCurr(\vx), \etaCurr, \vxCurr}$ where $\phiCurr : \RnToR$ and $\x_{k} \in \Rn$ are chosen according to some probability distribution is called a \emph{(probabilistic) estimate sequence of $f$} if $\lim_{k \rightarrow 0} \etaCurr = 0$  and for all $k \geq 0$ we have 
\begin{equation}
\forall \x \in \Rn 
\enspace : \enspace
\Eof{\phiCurr(\x)} \leq (1 - \eta_{k}) \f(\x) 
+ \eta_{k} \Eof{\phi_{0}(\x)}
\label{eq:lower_estimate_sequence2}
\end{equation}
and
\begin{equation}
\Eof{f(\vx_{k})} \leq \min_{\vx \in \Rn} \Eof{\phiCurr(\x)}
\label{eq:upper_estimate_sequence2}
\end{equation}
\end{defn}

A probabilistic estimation sequence gives a randomized minimization method due to the following 
\[
\lim_{k \rightarrow \infty} \Eof{f(\vxCurr)} - f^{*}
\leq \lim_{k \rightarrow \infty} \eta_{k}\left(\E\phi_{0}(x^{*})-f^{*}\right)
= 0
\enspace.
\]
Since a probabilistic estimation sequence can be constructed using random partial derivatives, rather than a full gradient computations, there is hope that in certain cases probabilistic estimation sequences require less information for fast convergence and therefore outperform their deterministic counterparts.

Similar to the accelerated gradient descent method, in the following lemma we first show how to combine a sequence of lower envelopes to satisfy condition (\ref{eq:lower_estimate_sequence2}) and prove that it preserves a particular structure on the current lower bound. 

\begin{lemma}[(Probabilistic) Estimate Sequence Construction]
\label{lem:prob_est_seq}
Let $\phiInit(\vx)$, $\infSeq{\vyCurr, \thetaCurr, \indexCurr}$ be such that
\begin{itemize}[noitemsep]
\item $\phi_{0} : \RnToR$ is an arbitrary function
\item Each $\vyCurr \in \Rn$
\item Each $\thetaCurr \in (0, 1)$ and $\sum_{k = 0}^{\infty} \thetaCurr = \infty$
\item Each $\indexCurr \in [n]$ is chosen randomly such that $\forall i \in [n]$ we have $\Pr[\indexCurr = i] = \frac{L_i^{\alpha}}{S_\alpha}$. 
\end{itemize}
Then the pair of sequences $\infSeq{\phiCurr(\vx), \etaCurr}$ defined by
\begin{itemize}
\item $\etaInit = 1$ and $\etaNext = (1 - \thetaCurr) \etaCurr$
\item $\phiNext(\vx) = (1 - \thetaCurr) \phiCurr(\vx) 
+ \thetaCurr \left[
\f(\vyCurr) 
+ \frac{S_{\alpha}}{L_{\indexCurr}} 
\innerprodOma{\gradfikVec(\vyCurr)}{\vx - \vyCurr}
+ \frac{\sigmaOma}{2} \normOma{\vx - \vyCurr}^2
\right]$
\end{itemize}
satisfies condition (\ref{eq:lower_estimate_sequence2}). Furthermore,
if $\phiInit(\vx) = \phiInit^{*} + \frac{\zetaInit}{2} 
\normOma{\vx - \vvInit}^{2}$,
then this process produces a sequence of quadratic functions of the
form
$
\phiCurr(\vx) = \phiCurr^{*} + \frac{\zetaCurr}{2}
\normOma{\vx - \vvCurr}^2
$
where
\begin{eqnarray}
\zetaNext 
& = & 
(1 - \thetaCurr) \zetaCurr + \thetaCurr \sigmaOma\label{eq:estimate_sequence_condition_1}
\\
v_{k+1} 
& = & 
\frac{1}{\zetaNext}\left[
(1 - \thetaCurr)\zetaCurr \vvCurr
+ \thetaCurr \sigmaOma \vyCurr 
- \frac{S_{\alpha} \thetaCurr}{L_{i_{k}}} \gradfikVec(\vyCurr) 
\right]
\label{eq:estimate_sequence_condition_2}
\\
\phiNext^{*} 
& = & 
(1-\thetaCurr )\phiCurr^{*}
+ \thetaCurr f(\vyCurr)
-\frac{\thetaCurr^{2} S_{\alpha}^{2}}{2\zetaNext} 
\frac{\left(f_{i_{k}}(\vyCurr)\right)^{2}}{L_{i_{k}}^{1+\alpha}}
\nonumber 
\\
&  & 
+
\frac{\thetaCurr(1 - \thetaCurr )\zetaCurr }{\zetaNext}
\left(\frac{\sigmaOma}{2} \normOma{\vyCurr-\vvCurr}^{2}
+\frac{S_{\alpha}}{L_{i_{k}}}
\innerprodOma{\vec{f}_{i_{k}}(\vyCurr)}{\vvCurr - \vyCurr}\right)
\enspace.\nonumber 
\end{eqnarray}
\end{lemma}

\begin{proof}
The proof follows from direct calculations however for completeness and for later use we provide a proof of an even more general statement in Appendix \ref{sec:appendix_prob_est_seq_form}.
\end{proof}

In the following theorem, we show how to choose $\vx$, $\vy$ and $\theta$
to satisfy the condition (\ref{eq:upper_estimate_sequence2}) and thereby derive a simple form of the general accelerated coordinate descent method. We also show that the number of iterations required for ACDM is 
$
\tilde{O}\left(\sqrt{\frac{S_{\alpha}n}{\sigma_{1-\alpha}}}\right)
$
which is strictly better than the number of iterations required for
coordinate descent method, $\tilde{O}\left(\frac{S_{\alpha}}{\sigma_{1-\alpha}}\right)$. Later in Theorem \ref{thm:lower_bound}, we show that this is optimal up to constant for the type of algorithm considered. Note that while several of the definitions specifications in the following theorem statement may at first glance seem unnatural our proof will show that they are nearly forced in order to achieve certain algorithm design goals.

\begin{theorem}[Simple ACDM]
\label{thm:simple_proof_acdm}
For all $i \in [n]$ let $\tilde{L}_i^\alpha \defeq \max(L_i^\alpha, S_\alpha /n)$ \footnote{We introduce this to avoid small sampling probabilities as they cause an issue in achieving the optimal $\tilde{O}\left(\sqrt{\frac{S_{\alpha}n}{\sigmaOma}}\right)$ convergence rate. This modification can be avoided by choosing a different distribution over coordinate updates in compute $x_{k + 1}$ which makes the algorithm more complicated and potentially more expensive.} and let $\tsa \defeq \sum_{i = 1}^{n} \tilde{L}_i^\alpha$.
Furthermore, for any $\vxInit \in \Rn$ and for all $k \geq 0$, let
\[
\vyInit = \vxInit
\qquad
\phiInit^* = f(\vxInit)
\qquad
\phi_0(\vx) = \phi_0^* + \frac{\sigmaOma}{2} \normOma{\vx - \vxInit}^2
\qquad
\thetaCurr = \sqrt{\frac{\sigmaOma}{2 \tSa n}}
\qquad
\]
Then applying Lemma \ref{lem:prob_est_seq} with these parameters, $\tilde{L}_i$ as the coordinate-wise gradient Lipshitz constants, and choosing $\vyCurr$ and $\vxCurr$ such that
\[
\forall k \geq 1
\enspace : \enspace
\frac{\thetaCurr \zetaCurr}{\zetaNext} (\vvCurr - \vyCurr)
+ \vxCurr - \vyCurr = 0
\enspace \text{ and } \enspace
\vxCurr = \vy_{k - 1} - \frac{1}{\tilde{L}_{i_{k - 1}}} \gradfikVec(\vy_{k - 1}) 
\]
yields a probabilistic estimate sequence. This \emph{accelerated coordinate descent method} satisfies
\begin{equation}
\label{eq:ACDM_simple_convergence}
\forall k \geq 0
\enspace : \enspace
\Eof{f(\vxCurr)} - f^{*}
\leq 
\left(1-\frac{1}{2}\sqrt{\frac{\sigma_{1-\alpha}}{S_\alpha n}}\right)^{k}
\left(f(\vxInit) - \optF + \sigmaOma \normOma{\optX - \vxInit}^{2}\right)
\enspace.
\end{equation}
\end{theorem}

\begin{proof}
By construction we know that condition \eqref{eq:lower_estimate_sequence2} of probabilistic estimation sequences holds. It remains to show that $\vxCurr$ satisfies condition \eqref{eq:upper_estimate_sequence2} and analyze the convergence rate. 
To prove \eqref{eq:upper_estimate_sequence2}, we proceed by induction to prove the following statement:
\[
\forall k \geq 0
\enspace : \enspace
\EcurrOf{f(\vxCurr)} \leq \EcurrOf{\min_{\vx \in \Rn} \phiCurr(\vx)}
= \EcurrOf{\phiCurr^*}
\enspace.
\]
where $\Ecurr$ indicates the expectation up to iteration $k$.
The base case $f(\vxInit) \leq \phiInit(\vxInit)$ is trivial and we proceed by induction assuming that $\EcurrOf{f(\vxCurr)} \leq \EcurrOf{\phiCurr^*}$. By Lemma \eqref{lem:prob_est_seq} and the inductive hypothesis we get
\begin{eqnarray*}
\EcurrOf{\phiNext^{*}}
& \geq & 
\E_{k}\left[
(1-\thetaCurr)f(\vxCurr) 
+ \thetaCurr f(\vyCurr) - \frac{\thetaCurr^{2}\tSa^{2}}{2\zetaNext}\frac{\left(f_{i_{k}}(\vyCurr)\right)^{2}}{\tLik^{1+\alpha}}
\right.
\\
 &  & 
 \left.
+ \frac{\thetaCurr (1-\thetaCurr )\zetaCurr }{\zetaNext}
\left(\frac{\sigmaOma}{2}
\normOma{\vyCurr - \vvCurr}^2
+\frac{\tSa}{\tLik}
\innerprodOma{\gradfikVec(\vyCurr)}{\vvCurr -\vyCurr}
\right)
\right]
\end{eqnarray*}
where for notational convenience we drop the expectation in each of the variables. By convexity $f(\vxCurr)\geq f(\vyCurr)+\left\langle \gradient f(\vyCurr),\vxCurr - \vyCurr\right\rangle $ so applying this and the definitions of $\normOma{\cdot}$ and $\innerprodOma{\cdot}{\cdot}$ we get
\[
\phiNext^{*}
\geq 
f(\vyCurr)-
\frac{\thetaCurr^{2}\tSa^{2} \gradfikVal(\vy_k)^2}{2\zetaNext \tLik^{1+\alpha}}
 +(1-\thetaCurr )\left(\frac{\tSa}{\tLik^{\alpha}}\left\langle \vec{f}_{i_{k}}(\vyCurr),\frac{\thetaCurr \zetaCurr }{\zetaNext}\left(\vvCurr -\vyCurr\right)\right\rangle +\left\langle \gradient f(\vyCurr),\vxCurr - \vyCurr\right\rangle \right)
\]
Using that $\forall i \in [n]$ we have $\Pr[i_k = i] = \frac{\tilde{L}_i^\alpha}{\tilde{S}_\alpha}$ we get
\[
\EnextOf{\phiNext^*}
\geq 
\EnextOf{f(\vyCurr)-\frac{\thetaCurr ^{2}\tSa^{2}}{2\zetaNext}\frac{\left(f_{i_{k}}(\vyCurr)\right)^{2}}{\tLik^{1+\alpha}}}
+
(1 - \thetaCurr ) \left\langle\grad f(\vyCurr),\frac{\thetaCurr \zetaCurr }{\zetaNext}\left(\vvCurr -\vyCurr\right) + \vxCurr - \vyCurr\right\rangle
\enspace.
\]
From this formula we see that $\vyCurr$ was chosen specifically to cancel the second term so that 
\[
\EnextOf{\phiNext^*}
\geq 
\EnextOf{f(\vyCurr)-\frac{\thetaCurr^{2}\tSa^{2}}{2\zetaNext}\frac{\left(f_{i_{k}}(\vyCurr)\right)^{2}}{\tLik^{1+\alpha}}}
= \sum_{i = 1}^{n}
\left[
f(\vyCurr) - 
\frac{\thetaCurr^{2} \tSa}{2\zetaNext} \frac{
\gradfiVal(\vyCurr)^{2}}{\tLi}
\right]
\]
and it simply remains to choose $\thetaCurr$ and $\vxNext$ so that $\EnextOf{f(\vxNext})$ is smaller than this quantity. 

To meet condition $\eqref{eq:upper_estimate_sequence2}$, we simply need to choose $\infSeq{\thetaCurr}$ so $\frac{\thetaCurr^2 \tSa}{\zetaNext} = \frac{1}{2 n}$. Using $\tilde{L}_i^\alpha \geq \frac{S_\alpha}{n} \geq \frac{\tsa}{2n}$, we have
\[
\EnextOf{\phiNext^*}
\geq \sum_{i = 1}^{n}
\left[
f(\vyCurr) - 
\frac{1}{2 \tSa} \frac{
\gradfiVal(\vyCurr)^{2}}{\tLi^{1-\alpha}}
\right]
\]
To compute $\vxNext$, we use the fact that applying Lemma \ref{lem:eqv_def_of_L} to the formula $f(\vy_{k} - t\gradfiVec(\vy_{k}))$ yields 
\[
f\left(\vy_{k}-\frac{1}{\tLi}\gradfiVec(\vy_{k})\right) 
\leq 
f(\vx) 
+ \left\langle \gradfiVec(\vy_{k}),-\frac{1}{\tLi}\gradfiVec(\vy_{k})\right\rangle 
+ \frac{\tLi}{2} \left(\frac{1}{\tLi}\gradfiVec(\vy_{k})\right)^{2}
\leq f(\vy_{k})-\frac{\gradfiVal(\vy_{k})^2}{2\tLi}
\label{eq:gradient_descent_jump}
\]
and therefore for $\vxNext$ as defined we have 
\[\EnextOf{f(\vxNext)} \leq \sum_{i = 1}^{n} \left[f(\vyCurr) - \frac{\gradfiVal(\vx)^2}{2 \tSa \tilde{L}_{i}^{1 - \alpha}}\right]
\leq  \EnextOf{\phiNext^*} 
\enspace.
\]
Recalling that
$
\zetaNext = (1 - \thetaCurr) \zetaCurr + \thetaCurr \sigmaOma
$
we see that choosing $\zetaInit = \sigmaOma$ implies $\zetaCurr = \sigmaOma$ for all $k$ and therefore choosing $\thetaCurr = \sqrt{\frac{\sigmaOma}{2 \tSa n}}$ completes the proof that the chosen parameters produce a probabilistic estimate sequence. Furthermore, we see that this choice implies that $\etaCurr = \left(1 - \sqrt{\frac{\sigmaOma}{2 \tSa n}}\right)^k$. Therefore, by the definition of a probabilistic estimate sequence and the fact that $\tSa \geq 2 S_\alpha$, equation \eqref{eq:ACDM_simple_convergence} follows.
\end{proof}

\subsection{Numerical Stability}
\label{sec:acdm_upper_num_stab}

In the previous section, we provided a simple proof how to achieve an ACDM with a convergence rate of $\tilde{O}(\sqrt{\frac{S_\alpha n}{\sigmaOma}})$. While sufficient for many purposes, the algorithm does not achieve the ideal dependence on initial error for certain regimes. For consistency with \cite{Nesterov2012} we perform the change of variables
\[
\forall k \geq 0
\enspace : \enspace
\alpha_{k} \defeq \frac{\thetaCurr \zetaCurr }{\zetaCurr +\thetaCurr \sigma_{1-\alpha}}
\qquad
\beta_{k} \defeq \frac{(1-\thetaCurr )\zetaCurr }{\zetaNext}
\qquad
\gamma_{k} \defeq \frac{\tilde{S_{\alpha}}\thetaCurr }{\zetaNext}
\]
and by better tuning $\thetaCurr$ we derive the following algorithm.

\begin{center}
\begin{tabular}{|l|}
\hline 
\textbf{Accelerated Coordinate Descent Method}\tabularnewline
\hline 
1. Define $\tilde{L_i} = \max(L_i,(S_\alpha /n)^{1/\alpha})$ and $\tilde{S}_\alpha = \sum \tilde{L}_i^\alpha$ \tabularnewline
\hline 
2. Define $\vvInit = \vxInit, a_{0}=\frac{1}{2n},b_{0}=2$\tabularnewline
\hline 
3. For $k\geq0$ iterate:\tabularnewline
\hline 
3a. Find $\alpha_{k},\beta_{k},\gamma_{k}\geq\frac{1}{2n}$ such that $\gamma_{k}^{2}-\frac{\gamma_{k}}{2n}=\left(1-\frac{\gamma_{k}\sigma_{1-\alpha}}{\tilde{S}_{\alpha}}\right)\frac{a_{k}^{2}}{b_{k}^{2}}=\beta_{k}\frac{a_{k}^{2}}{b_{k}^{2}}=\frac{\beta_{k}\gamma_{k}}{2n}\frac{1-\alpha_{k}}{\alpha_{k}}$.\tabularnewline
\hline 
3b. $\vyCurr=\alpha_{k}\vvCurr +(1-\alpha_{k}) \vxCurr.$ \tabularnewline
\hline 
3c. Choose $i_{k}$ according to $P_{\alpha}(i)=\tilde{L}_{i}^{\alpha} / \tilde{S_\alpha}$.\tabularnewline
\hline 
3d. $\vxNext =\vyCurr-\frac{1}{\tilde{L}_{i_{k}}}\vec{f}_{i_{k}}(\vyCurr),v_{k+1}=\beta_{k}\vvCurr +(1-\beta_{k})\vyCurr-\frac{\gamma_{k}}{\tilde{L}_{i_{k}}}\vec{f}_{i_{k}}(\vyCurr)$. \tabularnewline
\hline 
3e. $b_{k+1}=\frac{b_{k}}{\sqrt{\beta_{k}}}$ and $a_{k+1}=\gamma_{k}b_{k+1}$.\tabularnewline
\hline 
\end{tabular}
\par\end{center}

In Appendix \ref{sec:app:acdm_coefficients} and \ref{sec:app:num_stab}, we give a proof of convergence of this method in the unit-cost RAM model by studying the following potential function \cite{Nesterov2012} for suitable constants $a_{k}, b_{k} \in \R$,
\[
a_{k} \left(\Eof{f(\vxCurr)} - \optF\right) + b_{k} \Eof{\normOma{\vvCurr -x_{*}}^{2}}
\enspace.
\]
The following theorem states that error in vector updates does not grow too fast and Lemma \ref{lem:coeff} further states that errors in the coefficients $\alpha$, $\beta$, and $\gamma$ also does not grow too fast. Hence, $O(\log n)$ bits of precision is sufficient to implement ACDM with the following convergence guarantees. Below we state this result formally and we refer the reader to Appendix \ref{sec:app:num_stab} for the proof.

\begin{theorem}[Numerical Stability of ACDM]
\label{thm:main_result} 
Suppose that in each iteration of ACDM step 3d has additive error $\epsilon$, i.e. there exists $\verrOneCurr, \verrTwoCurr \in \Rn$ with $\normOma{\verrOneCurr} \leq \epsilon$ and $\normOma{\verrTwoCurr} \leq \epsilon$ such that step 3 is
\[
\vxNext := \vyCurr - \frac{1}{\tlik} \gradfikVec(\vyCurr) + \verrOneCurr
\enspace \text{ and } \enspace
\vvNext := \beta_{k} \vvCurr + (1 - \beta_{k}) \vyCurr - \frac{\gamma_{k}}{\tlik}
\gradfikVec(\vyCurr) + \verrTwoCurr.
\]
If $\epsilon < \frac{\sigmaOma^{2}}{8\tsa^{2}n}$ and $k\geq\sqrt{\frac{2 \tsa n}{\sigmaOma}}$, then we have the convergence guarantee
\begin{equation}
\sigmaOma 
\Eof{\normOma{\vvNext - \xopt}^2} + \left(\Eof{f(\vxNext)} - \optF \right) \leq\delta_{k}
\label{eq:ACD_bound}
\end{equation}
where 
\[
\delta_{k}\defeq24kS_{\alpha}\varepsilon^{2}+32\sigma_{1-\alpha}\left(1-\frac{1}{5}\sqrt{\frac{\sigma_{1-\alpha}}{S_{\alpha}n}}\right)^{k}\left(||x_{0}-x^{*}||_{1-\alpha}^{2}+\frac{1}{S_{\alpha}^{2}}\left(f(x_{0})-f^{*}\right)\right)
\]
and the additional convergence guarantee that
$
\frac{1}{k}\sum_{j=k}^{2k-1}
\normOma{\grad f(\vyCurr)}^{2} \leq 2000 \frac{\tsa}{n} \delta_{k}.
$
\end{theorem}

This theorem provides useful estimates for the error $f(\vxNext) - \optF$, the residual $\normOma{\vvNext - \optX}^{2}$, and the norm of gradient $\normOmaDual{\gradient f(\vyCurr)}$. Note how the estimate depends on the initial error $f(\vxInit) - \optF$ mildly as compared to Theorem \ref{thm:simple_proof_acdm}. We use this fact in creating an efficient SDD solver in \sectionref{sec:faster_sdd_solver}.

\subsection{Efficient Iteration}
\label{sec:efficient_iteration}

In both Nesterov's paper \cite{Nesterov2012} and later work \cite{Richtarik2011} the original ACDM proposed by Nesterov was not recommend since a naive implementation takes $O(n)$ time to update the vector $\vvCurr$, and therefore is likely slower than accelerated gradient descent. This implementation issue is a potential serious problem under the assumption that the oracle to compute gradient can only accept a single vector as input. However, if we make the mild assumption that we can compute $\gradient \f(t \vx + s \vy)$ for $s, t \in \R$ and $\vy, \vy \in \R$ in the same asymptotic runtime as it takes to compute $\gradient \f(\vx)$ (i.e. we do not need to compute the sum explicitly), then we can implement ACDM without additional asymptotic computational costs per iteration as compared to the cost of the coordinate descent method performing an update on the given coordinate. In the following lemma, we prove this fact and show that the technique can be executed in the unit-cost RAM model without additional asymptotic costs. Note that the method we propose does introduce potential numerical problems which we argue are minimal.

\begin{lemma}[Efficient ACDM Iteration]
\label{lem:stable}
For $S_{\alpha}=O(\poly(n))$ and $\sigma_{1-\alpha}=\Omega(\poly(\frac{1}{n}))$ each iteration of ACDM can be implemented in $O(1)$ time plus the time to make one oracle call of the form $\gradfikVec(t \vx + s \vy)$ for $s, t \in \R$ and $\vx, \vy \in \Rn$, using at most an additional $O(\log n)$ bits of precision so long as the number of iterations of ACDM is $O\left(\sqrt{\frac{S_\alpha n}{\sigmaOma}}\log(n)\right)$.
\end{lemma}
\begin{proof}
By the specification of the ACDM algorithm, we have
\begin{align*}
\vyNext
&= \alpha_{k + 1} \vvNext + (1 - \alpha_{k + 1}) \vxNext \\
&= \alpha_{k + 1} \left(\beta_{k} \vvCurr + (1 - \beta_k) \vyCurr + \frac{\gamma_k}{\tLik} \gradfikVec(\vy)\right)
+ (1 - \alpha_{k + 1}) \left(\vyCurr - \frac{1}{\tlik} \gradfikVec(\vyCurr)\right) \\
&= \alpha_{k + 1} \beta_{k} \vvCurr
+ \left(\alpha_{k + 1} (1 - \beta_{k}) + (1 - \alpha_{k + 1})\right) \vyCurr
- \frac{1 - \alpha_{k + 1} - \alpha_{k + 1} \gamma_k}{\tlik} \gradfikVec(\vy)
\end{align*}
we see that each update step of ACDM can be rewritten as
\[
\begin{pmatrix}
\vvNext^T \\
\vyNext^T
\end{pmatrix}
:=
\A_{k}
\begin{pmatrix}
\vvCurr^T \\
\vyCurr^T
\end{pmatrix}
- \vs_{k}
\enspace \text{ where } \enspace
\A_{k} = 
\begin{pmatrix}
\beta_{k} & 1-\beta_{k}\\
\alpha_{k+1}\beta_{k} & 1-\alpha_{k+1}\beta_{k}
\end{pmatrix}
\enspace \text{ and } \enspace
\vs_{k} = 
\begin{pmatrix}
\frac{\gamma_{k}}{\tlik}
\gradfikVec(\vyCurr)^T\\
\frac{1-\alpha_{k+1}-\alpha_{k+1}\gamma_{k}}{\tlik} 
\gradfikVec(\vyCurr)^T
\end{pmatrix}
\enspace.
\]
Therefore to implement ACDM in each iteration we can just maintain vectors $\vv, \vy \in \Rn$ and matrix $\mb_k \in \R^{2 \times 2}$ such that 
$
\begin{pmatrix}
\vvCurr^T \\
\vyCurr^T
\end{pmatrix}
= \mb_{k}
\begin{pmatrix}
\vv^T \\
\vy^T
\end{pmatrix}$
. With this representation each update step is
$
\mb_{k + 1}
\begin{pmatrix}
\vv^T \\
\vy^T
\end{pmatrix}
:=
\ma_{k} \mb_k
\begin{pmatrix}
\vv^T \\
\vy^T
\end{pmatrix}
- \vs_k
$
and by our oracle assumption, we can implement the following equivalent update step
\[
\mb_{k + 1} := \ma_{k} \mb_{k}
\enspace \text{ and } \enspace
\begin{pmatrix}
\vv^T\\
\vy^T
\end{pmatrix}
:=
\begin{pmatrix}
\vv^T\\
\vy^T
\end{pmatrix}
- 
\mb_{k + 1}^{-1} \vs_{k}
\]
in time $O(1)$ plus the time needed for one oracle call.

To complete the lemma, we simply need to show that this scheme is numerically stable. Note that 
\[
\det(\ma_{k})
= 
\beta_k (1 - \alpha_{k + 1} \beta_k) - \alpha_{k + 1} \beta_{k} (1 - \beta_k)
= \beta_k (1 - \alpha_{k + 1})
\enspace.
\]
Now by Lemma \ref{lem:coeff}, we know that 
$\alpha_{k} \leq 32 \max \left\{\frac{1}{k},\sqrt{\frac{\sigma_{1-\alpha}}{2 \tsa n}}\right\}$ 
for $k > 1$ and we know that 
$\beta_{k} \geq 1 - \frac{1}{2}\sqrt{\frac{\sigmaOma}{2\tsa n}}$ for all $k$. Therefore, letting $\kappa \defeq \sqrt{\frac{\sigma_{1-\alpha}}{2 \tsa n}}$ and recalling that we assumed that the total number of iterations is $O(\kappa^{-1} \log n)$ we get
\begin{eqnarray*}
\det(\mb_{k}) 
& \geq & 
\left(1-\kappa\right)^{O\left(\kappa^{-1} \log n\right)}(1-32\kappa)^{O(\kappa^{-1} \log n)}\prod_{k=2}^{\kappa^{-1}}\left(1-\frac{32}{k}\right)
\\
 & = &
\Omega\left(\poly\left(\frac{1}{n}\right)e^{-32\log\kappa}\right)
=
\Omega\left(\poly\left(\frac{1}{n}\right)\right).
\end{eqnarray*}
Hence, $O(\log n)$ bits of precision suffice to implement this method.
\end{proof}

\section{Faster Linear System Solvers}
\label{sec:applications}

In this section, we show how ACDM can be used to achieve asymptotic runtimes that outperform various state-of-the-art methods for solving linear systems in a variety of settings. In \sectionref{sec:conjugate_gradient}, we show how to use coordinate descent to outperform conjugate gradient under minor assumptions regarding the eigenvalues of the matrix under consideration, in \sectionref{sec:kaczmarz}, we show how to improve the convergence rate of randomized Kaczmarz type methods, and in \sectionref{sec:faster_sdd_solver}, we show how to use the ideas to achieve the current fastest known numerically stable solver for symmetric diagonally dominant (SDD) systems of equations.

\subsection{Comparison to Conjugate Gradient Method}
\label{sec:conjugate_gradient}

Here we compare the performance of ACDM to conjugate gradient (CG) and show that under mild assumptions about the linear system being solved ACDM achieves a better asymptotic running time.

For symmetric positive definite (SPD) matrix $\A \in \Rnn$ and vector $\vb \in \Rn$, we solve the linear system of equations $\A \x = \vb$ via the following equivalence unconstrained quadratic minimization problem:
\begin{equation}
\label{eq:pd_system}
\min_{\x \in \Rn}
\fun(\x)
\defeq
\frac{1}{2}
\innerprod{\A \x}{\x} - \innerprod{\vb}{\x}
\enspace.
\end{equation}

Let $m$ denote the number of nonzero entries in $\A$, let $\nnz_i$ denote the number of nonzero entries in the ith row of $\A$. To make the analysis simpler, we assume that the nonzero entries is somewhat uniform, namely $\forall i \in [n]$ we have $\nnz_i = O(\frac{m}{n})$. This assumption is trivially met for dense matrices, finite difference matrices, etc. Letting $0 \leq \lambda_1 \leq \ldots \leq \lambda_n$ denote the eigenvalues of $\A$, we get the following running time guarantee for ACDM.

\begin{theorem}[ACDM on SPD Systems]
Assume $\A$ is a SPD matrix with the property that $\nnz_i = O(\frac{m}{n})$ for all $i \in [n]$. Let the numerical rank $r(\A) = \sum_{i = 1}^{n} \lambda_i / \lambda_n$. ACDM applied to \eqref{eq:pd_system} with $\alpha = 1$\footnote{Here we only consider the $\alpha = 1$ case for easier comparison with conjugate gradient.} produces an approximate solution in 
$
\tilde{O}\left(m \sqrt{\frac{r(\A)}{n}} \sqrt{ \frac{\lambda_n}{\lambda_1}}\log \frac{1}{\epsilon} \right)
$
time with $\epsilon$ error in $\A$ norm in expectation.
\end{theorem}

\begin{proof}
The running time of ACDM with $\alpha = 1$ depends on $\sigma_0$ and $S_1$. From Example \ref{psd_example}, the total
component-wise Lipschitz constant $S_{1}$ is the trace of $A$, which
is $\sum\lambda_{i}$ and the convexity parameter $\sigma_{0}$ is
$\lambda_{1}$, therefore by Theorem \ref{thm:main_result} the convergence rate of ACDM is 
$
\sqrt{\frac{\lambda_{1}}{n\sum_{i=1}^{n}\lambda_{i}}}
$
as desired. Furthermore, the running time of each step depends on the running time of the oracle, i.e. computing
$
f_{i}(x)=\left(Ax\right)_{i}
$
, which by our assumption on $\nnz_i$ takes time $O\left(\frac{m}{n}\right)$.
\end{proof}

To compare with conjugate gradient, we know that one crude bound for the rate of convergence of conjugate gradient is $O\left(\sqrt{\frac{\lambda_{1}}{\lambda_{n}}}\right)$. Hence the total running time of CG to produce an epsilon approximate solution is 
$
\tilde{O}\left(m \sqrt{\frac{\lambda_{n}}{\lambda_{1}}} \log \frac{1}{\epsilon}\right) $. Therefore, with this bound ACDM is always faster or matches the running time since the numerical rank of $\A$ is always less than or equals to $n$. Thus, we see that when the numerical rank of $\A$ is $o(n)$, ACDM will likely \footnote{The running time of CG may be asymptotic faster when the eigenvalues form clusters.} have a faster asymptotic running time. 

To be more fair in our comparison to conjugate gradient, we note that in \cite{Spielman2009} tighter bound on the performance of CG was derived and they showed that in fact CG has a running time of 
$
\tilde{O}\left(\sum \nnz_{i}\left(\frac{\sum_{i=1}^{n}\lambda_{i}}{\lambda_{1}}\right)^{1/3}\right)
$
implying that ACDM is faster than CG when $\sum_{i=1}^{n}\lambda_{i}\leq n^{3}\lambda_{1}$ and it is usually satisfied. In the extreme cases that the condition is false, CG will need to run for $O(n)$ iterations at which point an exact answer could be computed. 
\subsection{Accelerating Randomized Kaczmarz}
\label{sec:kaczmarz}

The Kaczmarz method \cite{Kaczmarz} is an iterative algorithm to solve
$\A \x = b$ for any full row rank matrix
$\A \in \R^{m \times n}$. Letting $\vvar{a}_{i} \in \Rn$ denote the $i$-th row of the matrix $\A$, we know that the solution
of $\A \x = \vb$ is the intersection of the hyperplanes $H_{i} \defeq \{x:\left\langle \vvar{a}_{i}, \vx_{k}\right\rangle = \vb_{i}\}$. The Kaczmarz method simply iteratively picks one of these hyperplanes and projects onto it by the following formula:
\[
x_{k + 1} := 
\proj_{H_{i_{k}}}(\x_{k})
\enspace \text{ where } \enspace
\proj_{H_{i_{k}}}(\x_{k}) + \frac{\vb_{i_{k}} - \innerprod{a_{i_{k}}}{\x_{k}}}
{\norm{a_{i_{k}}}^{2}} a_{i_{k}}
\enspace.
\]

There are many schemes that can be chosen to pick the hyperplane $i_{k}$, many of which are difficult to analyze and compare, but in a breakthrough result, Strohmer and Vershynin in 2008 analyzed the randomized schemes which sample the hyperplane with probability proportional to $\normTwo{a_{i}}^{2}$. They proved the following

\begin{theorem}[Strohmer and Vershynin \cite{Strohmer2008}]
The Kaczmarz method samples row $i$ with probability proportionally to $\normTwo{a_i}^2$ at each iteration and yields the following
\[
\forall k \geq 0
\enspace : \enspace
\Eof{ \normTwo{\vxCurr - \xopt}^2 }
\leq
(1 - \kappa(\A)^{-2})^k \norm{\x_0 - \xopt}^2
\]
where $\optPoint \in \mathbb{R}^{n}$ is such that $\A \optPoint = \vb$, $\kappa(\mvar{A}) \defeq \normTwo{\A^{-1}} \cdot \normFro{\A}$ is the \emph{relative condition number} of $\A$, $\A^{-1}$ is the left inverse of $\A$, $\normTwo{\A^{-1}}$ is the smallest non-zero spectral value of $\A$ and $\norm{\A}_{F}^{2} \defeq \sum a_{ij}^{2}$ is the Frobenius norm of $\A$ 
\end{theorem}

Here we show show to cast this algorithm as an instance of coordinate descent and obtain an improved convergence rate by applying ACDM. We remark that accelerated Kaczmarz will sample rows with a slightly different probability distribution. As long as this does not increase the expected computational cost of an iteration, it will yield an algorithm with a faster asymptotic running time.

\begin{theorem}[Accelerated Kaczmarz]
The ACDM method samples row $i$ with probability proportionally to$\max \left\{\normTwo{a_i}^2, \frac{\normFro{\mvar{A}}^2}{m}\right\}$ and performs extra $O(1)$ work  at each iteration. It yields the following 
\[
\forall k \geq 0
\enspace : \enspace
\E \normTwo{\x_k - \xopt}^2
\leq 
3\left(1 - \frac{\kappa(\A)^{-1}}{2 \sqrt{m}}\right)^k 
\norm{\x_0 - \xopt}^2.
\]
\end{theorem}

\begin{proof}
To cast Strohmer and Vershynin's randomized Kaczmarz algorithm in the framework of coordinate descent, we consider minimizing the objective function of theorem theorem directly, i.e.
$
\min_{\x \in \Rn} 
\frac{1}{2}
\normTwo{\x - \optPoint}^2
$.
Since $\A$ has full row rank, we write $\x = \A^{T} \y$ and consider the equivalent problem
$
\min_{\y \in \Rm}
\frac{1}{2} \normTwo{\A^{T} \y - \optPoint}^{2}
$.
Expanding the objective function and using $\A \x^* = \vb$, we get
\[
\normTwo{\A^{T} \y - \optPoint}^{2}
=
\normTwo{\A^{T} \y}^{2}
-2 \innerprod{\vb}{\y}
+ \normTwo{\optPoint}^{2}
\enspace.
\]
Therefore, we attempt solve the following equivalent problem using accelerated coordinate descent.
\[
\min_{\y \in \Rm}
f(\y)
\enspace \text{ where } \enspace
f(\y) \defeq  \frac{1}{2}
\normTwo{\A^{T} \y}^{2} - 
\innerprod{\vb}{\y}
\enspace.
\]
From Example \ref{linear_system_example}, we know that the $i$-th direction component-wise Lipschitz constant is $L_{i}= \norm{a_{i}}^{2}$ where $a_{i}$ is the $i$-th row of $\A$ and we know that $\grad f(\vy) = \A \A^T \vy - b$. Therefore, each step of the coordinate descent method consists of the following \footnote{We ignore the thresholding here for illustration purpose.}
\[
\vyNext
\enspace :=  \enspace
\vyCurr - \frac{1}{L_{i_{k}}} \gradfikVec(\vyCurr)
\enspace = \enspace 
\vyCurr - \frac{1}{\norm{\vvar{a}_{i_{k}}}^{2}}\left(\A \A^{T} \vyCurr - \vb\right)_{i_{k}}
\enspace.
\]
Recalling that we had performed the transformation $\vx = \A^T \y$ we see that the corresponding step in $\x$ is 
\[
\vxNext
\enspace  = \enspace
\vxCurr - \frac{1}{\norm{a_{i}}^{2}}
\left(\A \x_{k} - \vb\right)_{i_{k}}
\enspace = \enspace
\vxCurr + 
\frac{\vb_{i_{k}} - \innerprod{\vvar{a}_{i_{k}}}{\vxCurr}}
{\normTwo{a_{i_{k}}}^{2}} \vvar{a}_{i_{k}}.
\]
Therefore, the randomized coordinate descent method applied this way yields precisely the randomized Kaczmarz method of Strohmer and Vershynin.

However, to apply the ACDM method and provide complete theoretical guarantees we need to address the problem that $f$ as we have constructed it is not strongly convex. This is clear by the fact that the null space of $\A^{T}$ may be non
trivial. 

To remedy this problem we let $Z \subseteq \Rm$ denote the null space of $\A^{T}$, i.e. $Z \defeq \{\vx \in \Rm ~ | ~ \A^T \vx = 0\}$, and we define the semi-norm $\norm{\cdot}_{Z^{\perp}}$ on $\Rm$ by
$
\norm{\vy}_{Z^{\perp}}
\defeq 
\inf_{\vvar{z} \in Z} \norm{y+z}
$. Now it is not hard to see that $f$ is strongly convex under this seminorm  with convexity parameter $\sigma_{Z^\perp} = \norm{\A^{-1}}_2^{-2}$. Furthermore, one can prove similarly to the proof of Lemma \ref{lem:prob_est_seq} and Theorem \ref{thm:simple_proof_acdm} that the algorithm in this theorem achieves the desired convergence rate.

To see this, we let $\Pmat \in \R^{m \times m}$ denote the orthogonal projection onto the image of $\A$ we note that $\norm{\vy}_{Z^{\perp}} = \sqrt{\vy^T \Pmat \vy}$. Therefore we can apply the general form of Lemma \ref{lem:prob_est_seq} proved in Appendix \ref{sec:appendix_prob_est_seq_form} to derive a similar estimate sequence result for this norm. Unfortunately, the resulting estimate sequence requires working with $\pseudo{\Pmat} \gradfikVal(\vyCurr)$. However, the proof of Theorem \ref{thm:simple_proof_acdm} still works using the same algorithm because of the facts that $\pseudo{\Pmat} \grad f(\vx) = \grad f(\vx)$, 
\[
\forall \vy \in \Rn, \forall \vz \in Z
\enspace : \enspace
f(\vy + \vz) = f(\vy)
\enspace \text{ and } \enspace
\grad f(\vy + \vz) = \grad(\vy)
\enspace,
\]
and $\norm{\vx}_2^2 \geq \norm{\vx}_{Z^\perp}^2$. This gives the same convergence guarantee with $\sigma_{1 - \alpha}$ replaced with $\norm{\A^{-1}}_2^{-2}$, $S_{\alpha} = \normFro{\A}^2$ and $\normOma{\cdot} = \norm{\cdot}_{Z^\perp}$. The desired convergence rate follows from the strong convexity of $f$. 

To justify the cost of each iteration, note that although the iterations work on the $\y$ variable, we can apply $\A^{T}$ on all iterations and just do the operations on $\x$ variables which is more efficient.
\end{proof} 
\subsection{Faster SDD Solvers in the Unit-cost RAM Model}
\label{sec:faster_sdd_solver}

A matrix $\A \in \Rnn$ is called \emph{Symmetric Diagonally Dominant (SDD)} if
$\A^T = \A$ and $\forall i \neq j \in [n]$ we have $\A_{ii} \geq \sum_{i \neq j} |A_{ij}|$. Solving such systems has had numerous applications in both theoretical and applied computer science. For an overview of such systems of equations, their applications, and their solvers we refer the reader to \cite{Spielman:2012:AGT:2359888.2359901}.

The fastest running times in general for solving such systems are $\tilde{O}(m \log n \log \frac{1}{\varepsilon})$\footnote{Where we use $\tilde{O}$ to hide $O(\log \log n)$ terms.} due to Koutis, Miller, and Peng (cite KMP). However, the numerical stability of this algorithm is difficult to bound and when analyzed in the standard unit-cost RAM model, the best known running time is $\tilde{O}(m \log^2 n \log \frac{1}{\varepsilon})$ due to Kelner \emph{et al.} \cite{Kelner2013}. 

Here we show how to cast the simplest solver presented \cite{Kelner2013} as an instance of coordinate descent, and by applying ACDM, we obtain a faster running time of $\tilde{O}(m \log^{3/2} n \log \frac{1}{\varepsilon})$ in the unit-cost RAM model. Our presentation will make heavy use of several insights from \cite{Kelner2013} and we refer the reader to that paper for further background.

Following the reasoning in \cite{Kelner2013}, solving SDD systems can be reduced to solving the Laplacian system $\lap \vx = \demands$ corresponding to a weighted connected graph $G = (V, E, \omega)$ where we call $\omega_e$ the weight of edge $e \in E$ and $r_e \defeq \frac{1}{\omega_e}$ the resistance. For notational convenience, we arbitrarily orient each edge in $E$ and using this convention define a graph's incidence matrix $\incMatrix \in \Rev$, resistance matrix $\rMatrix \in \R^{E\times E}$, and Laplacian matrix, $\lap \in \R^{V \times V}$ as follows:
\[
\forall (a, b), e_1, e_2 \in E
\enspace : \enspace
\incMatrix_{(a,b),c} \defeq \indicVec{a=c} - \indicVec{b = c},
\quad 
\rMatrix_{e_{1},e_{2}} \defeq r_{e_1} \cdot \indicVec{e_{1} = e_{2}},
\quad 
\lap \defeq \incMatrix^{T} \rMatrix^{-1} \incMatrix.
\]
Letting $\vx^*$ be the solution of $\lap \vx = \demands$, we will prove the following:

\begin{theorem}
By applying ACDM to a simple Laplacian system solver in \cite{Kelner2013}, we can produce an $\vx \in \Rn$ such that
$
\normL{\vx - \vx^*} \leq \epsilon \normL{\vx^*}
$
in time $\sddRuntimeOurs$.\footnote{Although not the focus of this section, we remark that this procedure produces an $\epsilon$-approximate electric flow (that is a $\vvar{y}$ such that $\incMatrix^T \vy = \demands$ and has an objective function value within a multiplicative $(1 + \epsilon)$ of the optimum) in time $\sddRuntimeOursFlow$ in the unit-cost RAM model. Furthermore, by the defintion of ACDM, we see that the operators produced by this algorithm are linear.} 
\end{theorem}

\begin{proof}
By the method of Lagrange multipliers, the minimizer $\vz^*$ of the problem
\[
\min_{\incMatrix^{T} \vz = \demands}\frac{1}{2}
\normR{\vz}^{2}
\enspace \text{ where } \enspace
\normR{\vz} \defeq \sqrt{\sum_{e \in E} r_e \vz_e^2}
\]
is $\rMatrix^{-1} \incMatrix \xopt$. Therefore, we can find $\xopt$ by minimizing $f(\vz) \defeq \frac{1}{2} \normR{\vz}^2$ and use a BFS to get $\vx$ from $\vz$. 

Now, let $\vz_0$ be any vector such that $\incMatrix^{T} \vz_0 = \demands$. With this, the problem can be simplified to
\[
\min_{\incMatrix^{T} \vy = 0} \frac{1}{2} \normR{\vz_{0} + \vy}^{2}.
\]
Now, the $\{\vc \in \Redgevec | \incMatrix^T \vc = 0 \}$ is simply the set of circulations of the graph, called \emph{cycle space}, and therefore to turn this constrained problem into an unconstrained problem we simply require a good basis for cycle space.

Such a basis can easily be found by a spanning tree. Given a spanning
tree $\tree$ of $G$, for any $(a,b) \in \offtreeEdgeSet$ let $\treeCycleVec{(a, b)} \in \Redgevec$ be the circulation that corresponds to sending 1 unit on $(a, b)$ and $1$ unit on the unique path from $b$ to $a$ in $T$. Now, the set $\{\treeCycleVec{e} ~ | ~ e \in \offtreeEdgeSet\}$ forms a basis for cycle space and with this insight we can simplify the problem further to 
\[
\min_{\vy \in \Roffedgevec} \frac{1}{2}
\normR{\vz_{0} + \cycleMatrix \vy}^{2}
\enspace \text{where} \enspace
\cycleMatrix \defeq
\left[c_{e_{1}} c_{e_{2}}\cdots\right] \in \R^{E \times E \setminus \tree}
\enspace.
\]
Now, for every \emph{off-tree edge}, i.e. $e \in \offtreeEdgeSet$, there is only one cycle $c_{e}$ that passes through it. So, if we let $\rMatrix_{\offtreeEdgeSet}$ be the diagonal matrix for the resistances of the off tree edges we have that for any $\vy \in \offtreeEdgeSet$
\[
\vy^T \cycleMatrix^T \rMatrix \cycleMatrix \vy
\geq \vy^T \cycleMatrix^T \rMatrix_{\offtreeEdgeSet} \cycleMatrix \vy
= \sum_{e \in \offtreeEdgeSet} r_e \vy(e)^2
\geq \left(\min_{e \in \offtreeEdgeSet} r_e\right) \normTwo{\vy}^2 .
\]
Therefore, the convexity parameter of $\normR{\vz_{0} + \cycleMatrix \vy}^{2}$ is at least $\min_{e\in E\backslash T} r_e$. This could be wasteful if the resistances vary, so we compensate by rescaling the space, $\vytilde = \rMatrix^{1/2} \vy$, to get the following problem:
\[
\min_{\vytilde \in \Roffedgevec} \fun(\vytilde)
\enspace \text{ where } \enspace
\fun(\vytilde) = \frac{1}{2} \normR{\vz_0 + \cycleMatrix \rMatrix^{-1/2} \vytilde}^{2} .
\]
By the same reasoning as above, the convexity parameter of $f$ with respect to the Euclidian norm is $1$ and we bound the $e$-th direction component-wise Lipschitz constant $L_{e}$ as follows
\[
\forall e \in \offtreeEdgeSet
\enspace : \enspace
L_e
= \indicVec{e}^T \rMatrix^{-1/2} \cycleMatrix^T \rMatrix \cycleMatrix \rMatrix^{-1/2} \indicVec{e}
= \frac{\treeCycleVec{e}^T \rMatrix \treeCycleVec{e}}{r_e}
= \stretchEdge{e} + 1 .
\]
where $\stretchEdge{e}$ denotes the \emph{stretch of $e$ by the tree $\tree$}, i.e. the sum of the resistances of the edges on the path in $\tree$ connecting the endpoints of $e$ divided by the resistance of the edge $e$. Therefore, the total component-wise Lipschitz constant is given by
\[
S_1 = \sum_{e \in \offtreeEdgeSet} (\stretchEdge{e} + 1)
= m - n + 1 - n + 1 + \sum_{e \in E} \stretchEdge{e}
\leq m + \sum_{e \in E} \stretchEdge{e} .
\]
Now, as in \cite{Kelner2013}, we can using the following result of Abraham and Neiman to ensure $S_1 = O(m \log n \log \log n)$
\begin{theorem}
\cite{Abraham2012} In $O(m\log n\log\log n)$ time, we can compute a
spanning tree $T$ such that 
\[
\sum_{e \in E} \st(e) = O(m\log n\log\log n).
\]
\end{theorem}
Now, applying the coordinate descent method to $\fun$ and letting $e_k \in \offtreeEdgeSet$ denote the off-tree edge i.e., coordinate, picked in iteration $k$, we get \footnote{We ignore the thresholding here again for illustration purpose.}
\begin{align*}
\vytilde_{k+1} 
& := 
\vytilde_{k} - \frac{1}{L_{e_{k}}} \vec{f}_{e_{k}}(\vytilde_{k})
\tag{Definition of Update Step}
\\
& = 
\vytilde_{k} - \frac{1}{L_{e_{k}}}
\left(
\rMatrix^{-1/2} \cycleMatrix^{T} \rMatrix (\vz_0 + \cycleMatrix \rMatrix^{-1/2} \vytilde)
\right)_{e_{k}}
\cdot
\indicVec{e_{k}}
\tag{Computation of $\grad f(\vytilde)$}
\\
& =
\vytilde_{k} - 
\frac{1}{L_{e_{k}} r_{e_{k}}^{1/2}}
\sum_{e} \treeCycleVec{e_k}(e) r_{e_k} (\vz_0 + \cycleMatrix \rMatrix^{-1/2} \vytilde)(e)
\cdot
\indicVec{e_{k}}
\tag{Definition of $\cycleMatrix$ and $\rMatrix$}
\\
& =
\vytilde_{k} - 
\frac{1}{L_{e_{k}} r_{e_k}^{1/2}}
\sum_{e \in \treeCycleVec{e_k}} r_e (\vz_0 + \cycleMatrix \rMatrix^{-1/2} \vytilde)(e)
\cdot
\indicVec{e_{k}}
\tag{Definition of $\treeCycle{e_k}$}
\end{align*}
Recalling $\vy = \rMatrix^{-1/2} \vytilde$ and the derivation of $L_e$, we can write this equivalently as
\[
\vy_{k + 1}
:= 
\vy_k - \left[\frac{1}{r_e(\st(e_k) + 1)} \sum_{e \in \treeCycleVec{e_k}} r_e (\vz_0 + \cycleMatrix \vy)(e)\right] \cdot \indicVec{e_k} .
\]
Noting that the $\vz$ with $\incMatrix^T \vz = \demands$ corresponding to $\vy$ is $\vz = \vz_0 + \cycleMatrix \vy$, we write the update equivalently as
\[
\vz_{k + 1}
:=
\vz_k
-
\left[
\frac{1}{r_e (\st(e_k) + 1)} \sum_{e \in \treeCycleVec{e_k}} r_e \cdot \vz_k(e)
\right] \cdot \treeCycleVec{e_k}
\]
which is precisely the algorithm of the simple solver in \cite{Kelner2013}. In \cite{Kelner2013}, they also prove that calls to $\vec{f}_{e_{k}}$ and updates to $\vy_{e_k}$ can be implemented in $O(\log n)$. Therefore, by applying ACDM we can obtain a faster algorithm. Note that apply ACDM efficiently computations of $\vec{f}_{e_k}$ need to be performed on the sum of two vector without explicit summing them. However, since in this case $\gradient f$ is linear, we can just call the oracle on the two vectors separately and also use the data structure for updating coordinates in each vector separately.

While the above insight suffices to compute \emph{electric flows}, i.e. $\vz$, in the desired runtime in order to actually solve the Laplacian system we need to be able to compute $\vx$ such that the following holds. 
\begin{equation}
\label{eq:Laplacian_sol}
\normL{\vx - \lapPseudo \demands}^{2}
\leq
\varepsilon \normL{\lapPseudo\demands}^{2}
\enspace.
\end{equation}
However, by Lemma 6.2 of \cite{Kelner2013}, it suffices to show that $\normTwo{\grad f(\vytilde)}^2 \leq \epsilon \optValue$. Note that
\[
\normTwo{\gradient \f(\vytilde)}^{2} 
= 
\normTwo{\rMatrix^{-1/2} \cycleMatrix^T \rMatrix (\vz_0 + \cycleMatrix \rMatrix^{-1/2} \vytilde)}^2
=
\sum_{e \in E \setminus \tree}
\frac{1}{r_e} \left(\sum_{e' \in \treeCycleVec{e}} \vz(e') r_{e'}\right)^2
\]
Now, suppose we choose $\vytilde_0 = \vvar{0}$. Then, we see that 
$
\normTwo{\vytilde_0 - \vytilde^*}^2 
= \normTwo{\vytilde^*}^2 
= \normTwo{\rMatrix_{E \setminus \tree}^{1/2} \vy^*}^2
\leq 2 \optValue,
$
and using Lemma 6.1 of \cite{Kelner2013}, we have that 
$\frac{1}{S_\alpha^2} (f(\vytilde_0) - f^*) \leq \frac{f^*}{S_\alpha} \leq f^*$. Furthermore, since by our choice of spanning tree $\frac{S_1}{|\offtreeEdgeSet|} = O(\log n \log \log n)$, after $k = O(m \sqrt{\log n \log \log n} \log \frac{\log n}{\epsilon})$ iterations of ACDM, by Theorem \ref{thm:main_result}, we have that
$
\frac{1}{k} \sum_{j = k}^{2k -1}
\normTwo{\gradient f(\vy_k)}^2 \leq \epsilon \optValue .
$
Therefore, if we stop ACDM at random iteration between $k$ and $2k - 1$, we have that $\E \normTwo{\gradient f(\vy_k)}^2 \leq \epsilon f^*$ as desired. Therefore, in $O(m\log^{3/2}n\sqrt{\log\log n}\log(\frac{\log n}{\varepsilon}))$
time, we can compute $\x$ satisfying (\ref{eq:Laplacian_sol}). 
\end{proof}

Beyond obtaining a faster running time, we remark that this algorithm did not require recursive techniques that were necessary for Kelner \emph{et al.} to achieve there fastest running time. The simple solver that was accelerated actually had a running time of $O(m\log^{2}n\log\log n\log(n/\varepsilon))$ time due to the potentially large initial error of $f(\vytilde_{0})$. To achieve their fastest running time, they ran their solver several times to ultimately remove the $n$ term in $\log(n/\varepsilon)$. In our case, since ACDM measures initial error both in terms of $f(\vytilde_0) - f^*$ and $\norm{\vytilde - \vytilde^*}^2$ and since $\norm{\vytilde - \vytilde^*}^2$ is relatively small, ACDM avoids this issue entirely.
\section{Towards the Optimality of Accelerated Coordinate Descent}
\label{sec:acdm_lower}

In this section, we prove that the accelerated coordinate
descent is optimal under the assumption that the iterative method generates a sequence of vectors $\{x_{k}\}\in\mathbb{R}^{n}$
such that 
\begin{equation}
\vxNext \in \vxInit + \text{span}\left\{ \vec{f_{i_{0}}}(\vx_{0}), \vec{f_{i_{1}}}(\vx_{1}),\cdots, \vec{f_{i_{k}}}(\vx_{k})\right\} .\label{eq:lower_bound_assumption}
\end{equation}
This assumption forbids the iterative method from starting at any other point
other than the initial point and forbids the algorithm from randomly jumping to completely new points. We do not know if this assumption is necessary.
However, if the iterative method can only access a function value
$f(\vx)$ and a derivative $f_{i}'(\vx)$ in a random direction at each
iteration, then at least $O(n^{2})$ iterations are needed to obtain full local second order information, i.e. $\hessian \f$. Therefore maybe this assumption is not needed. 
\begin{theorem}[ACDM Lower Bound]
\label{thm:lower_bound}
Assume the iterative method satisfies the assumption (\ref{eq:lower_bound_assumption}) and further assume the method randomly picks each $i_k$ uniformly at random.\footnote{We make this assumption to simplify analysis. However we remark that we could repeat our analysis by the coordinates randomly permuted. Then by Yao's minimax principle one could quantify how much better an algorithm could do on this hard distribution of functions.
} Then, for any $S_{1}>4\sigma n>0$, $x_{0}\in\mathbb{R}^{n}$ and $k\leq\frac{n^{2}}{2}$,
there exists a convex function $f$ with strong convex parameter $\sigma$
and total component-wise Lipschitz constant $S_{1}$ such that 
\begin{eqnarray*}
\Eof{f(\vxCurr)} - f^*
& \geq & 
\frac{\sigma}{2}\left(1 - 2\sqrt{\frac{2\sigma}{nS_{1}}}\right)^{k}
\normTwo{\xopt - \vxInit}^{2} 
-\sqrt{\frac{\sigma S_{1}}{n}}
\left(1-\frac{1}{2}\sqrt{\frac{n\sigma}{S_{1}}}\right)^{2n}
\\
 & \sim & 
 \frac{\sigma}{2}\left(1 - 2\sqrt{\frac{2\sigma}{nS_{1}}}\right)^{k}
 \normTwo{\xopt - \vxInit}^{2}
 \enspace.
\end{eqnarray*}
\end{theorem}
\begin{proof}
We assume without loss of generality that $\vx_{0} = \vzero$ by shifting the domain and let 
\[
f(\vx)
\defeq
\frac{L-\sigma}{4}
\left(
(1-\vx(1))^{2}
+\sum_{i=1}^{n-1}(\vx(i)-\vx(i+1))^{2}
+\left(\vx(n)-q^{n+1}\right)^{2}
\right) 
+
\frac{\sigma}{2}\sum_{i=1}^{n}\vx^{2}(i)
\]
where $L \defeq S_{1}/n$ and $q$ is a constant to be determined later. It is
easy to check that $f$ is $\sigma$ strongly convex with respect to $\normTwo{\cdot}$ and that the total component-wise Lipschitz constant is $L$. Taking the derivative of $\vx(i)$ we see that the minimizer of $f$, denoted $\xopt$ satisfies
\[
\forall k
\enspace : \enspace
\frac{L-\sigma}{2}\left(\xopt(k) - \xopt(k-1) + \xopt(k) - \xopt(k+1)\right) +\sigma \xopt(k)=0
\]
where we define $\xopt(0) \defeq 1$ and $\xopt(n + 1) \defeq q^{n + 1}$. From this we see that
\[
\xopt(k + 1)-\frac{2L}{L-\sigma}\xopt(k) + \xopt(k - 1)=0
\enspace.
\]
and $\xopt(k) = C_{1}(q^{+})^{k}+C_{2}(q^{-})^{k}$ where $q^{\pm}=\frac{L}{L-\sigma}\pm\sqrt{(\frac{L}{L-\sigma})^{2}-1}$
and $C_{1},C_{2}$ are some constants that depend on the first and the
last terms. To simply the term, we set $q=q^{-}$ and get
\[
x^{*}(k)=q^{k}.
\]
Now, we want to show that the $\vxCurr$ produced by the iterative method
is likely far away from optimal since $\vxCurr$ likely has many zeros when
$k$ is small. We say $x \in R^{j}$ if only the first $j$
coordinates of $\vx$ are nonzero. If $x \in R^j$ by the strong convexity of $f$ we have
\begin{equation}
f(\vx)-f(\xopt) 
\geq \frac{\sigma}{2}\normTwo{\vx - \xopt}^{2}
\geq \frac{\sigma}{2}\sum_{i=j+1}^{n}q^{2i}
\geq \frac{\sigma}{2}q^{2j} \normTwo{\xopt}^{2}-\frac{\sigma}{2}\frac{q^{2n+2}}{1-q^{2}}.\label{eq:lower_bound_f_if_many_zero}
\end{equation}
Note that $\vxInit \in R^0$ and if $\vx \in R^{j}$ then $
f_{k}(\vx) = 0$ for $k > j + 1$. Furthermore, by the assumption (\ref{eq:lower_bound_assumption}), the current point $\vx$ in an iteration goes from $R^{k}$ to $R^{k+1}$ only if the oracle gives the $(k+1)$-th partial derivative. Since by assumption this happens with probability $\frac{1}{n}$ by applying \eqref{eq:lower_bound_f_if_many_zero}, we have
\begin{eqnarray*}
Ef(x_{k})-f(x^{*}) 
& \geq & 
\sum_{i=0}^{k}
\binom{k}{i}\left(\frac{1}{n}\right)^{i}\left(1-\frac{1}{n}\right)^{k-i}\left(\frac{\sigma}{2}q^{2i}
\normTwo{\xopt}^{2}
-\frac{\sigma}{2}\frac{q^{2n+2}}{1-q^{2}}\right)\\
 & = & \frac{\sigma}{2}\left(1-\frac{1}{n}+\frac{q^{2}}{n}\right)^{k}
 \normTwo{\xopt}^{2}-\frac{\sigma}{2}\frac{q^{2n+2}}{1-q^{2}}.
\end{eqnarray*}
where the last line comes from the fact that $E(t^{X})=(1-p+pt)^{n}$
if $X$ is a Binomial distribution with probability $p$ with $n$
trial. 

Next by taking the Taylor series at $\frac{L}{L-\sigma}=1$ and use the assumption
$S_{1}>4 \sigma n$, we have
\[
1-\sqrt{\frac{2\sigma}{L}}\leq q\leq1-\frac{1}{2}\sqrt{\frac{\sigma}{L}}.
\]
 Therefore, we have 
\begin{eqnarray*}
Ef(x_{k})-f(x^{*}) 
& \geq & 
\frac{\sigma}{2}\left(1-\frac{1}{n}+
\frac{\left(1-\sqrt{\frac{2\sigma}{L}}\right)^{2}}{n}\right)^{k}
\normTwo{\xopt}^{2}-\frac{\sigma}{2}\frac{q^{2n+2}}{1-q^{2}}\\
 & \geq & 
 \frac{\sigma}{2}\left(1-\frac{2}{n}\sqrt{\frac{2\sigma}{L}}\right)^{k}
 \normTwo{\xopt}^{2}-\frac{\sigma}{2(1-q^{2})}\left(1-\frac{1}{2}\sqrt{\frac{\sigma}{L}}\right)^{2n+2}\\
 & \geq & 
 \frac{\sigma}{2}\left(1-\frac{2}{n}\sqrt{\frac{2\sigma}{L}}\right)^{k}
 \normTwo{\xopt}^{2}-\sqrt{\sigma L}\left(1-\frac{1}{2}\sqrt{\frac{\sigma}{L}}\right)^{2n}.
\end{eqnarray*}
\end{proof}

\section{Acknowledgements}

We thank Yan Kit Chim, Andreea Gane, Jonathan Kelner, Lap Chi Lau, and Lorenzo Orecchia for helpful discussions. This work was partially supported by NSF
awards 0843915 and 1111109 and a NSF Graduate Research Fellowship (grant no. 1122374).

\newpage

\bibliographystyle{plain}
\bibliography{lsFOC13}

\appendix

\section{Proof of Probabilistic Estimate Sequence Form}
\label{sec:appendix_prob_est_seq_form}

Here we prove a stronger variant of Lemma \ref{lem:prob_est_seq} that is useful for Section \ref{sec:kaczmarz}. Throughout this section we let $\A \in \Rnn$ denote a symmetric positive semidefinite matrix, we let $\normA{\vx} \defeq \sqrt{\vx^T \A \vx}$ be the induced norm, we let $\innerprodA{\vx}{\vy} \defeq \vx^T \A \vy$ be the inner product which induces this norm, and we let $\pseudo{\A}$ denote the Moore-penrose pseudoinverse of $\A$. Taking $\A$ to be the diagonal matrix with $\A_{ii} = L_{1 - \alpha}$ and applying the following lemma yields Lemma \ref{lem:prob_est_seq} as an immediate corollary.

\begin{lemma}[General (Probabilistic) Estimate Sequence Construction]
Let $\A$ be a positive semidefine matrix such that $\f$ is strongly convex with respect to the norm $\norm{\cdot}_{\A}$ with convexity parameter $\sigma_{\A}$. Furthermore let $\phiInit(\vx)$, $\infSeq{\vyCurr, \thetaCurr, \indexCurr}$ be such that
\begin{itemize}[noitemsep]
\item $\phi_{0} : \RnToR$ is an arbitrary function
\item Each $\vyCurr \in \Rn$,
\item Each $\thetaCurr \in (0, 1)$ and $\sum_{k = 0}^{\infty} \thetaCurr = \infty$,
\item Each $\indexCurr \in [n]$ is chosen randomly with probability $\probCurr$ .
\end{itemize}
Then the pair of sequences $\infSeq{\phiCurr(\vx), \etaCurr}$ defined by
\begin{itemize}
\item $\etaInit = 1$ and $\etaNext = (1 - \thetaCurr) \etaCurr$,
\item $\phiNext(\vx) := (1 - \thetaCurr) \phiCurr(\vx) ,
+ \thetaCurr \left[
\f(\vyCurr) 
+ \frac{1}{p_{i_k}}
\innerprod{\gradfikVec(\vyCurr)}{\vx - \vyCurr}
+ \frac{\sigmaA}{2} \normA{\vx - \vyCurr}^2
\right]$
\end{itemize}
satisfies condition (\ref{eq:lower_estimate_sequence2}). Furthermore,
if $\phiInit(\vx) = \phiInit^{*} + \frac{\zeta_{0}}{2} 
\normA{\vx - \vvInit}^{2}$,
then this process produces a sequence of quadratic functions of the
form
$
\phiCurr(\vx) = \phiCurr^{*} + \frac{\zetaCurr}{2}
\normA{\vx - \vvCurr}^2
$
where
\begin{eqnarray}
\zetaNext 
& = & 
(1 - \thetaCurr) \zetaCurr + \thetaCurr \sigmaA,
\nonumber
\\
\vvNext 
& = & 
\frac{1}{\zetaNext}\left[
(1 - \thetaCurr)\zetaCurr \vvCurr
+ \thetaCurr \sigmaA \vyCurr 
- \frac{\thetaCurr}{\probCurr} \pseudo{\A} \gradfikVec(\vyCurr) 
\right],
\nonumber 
\\
\phiNext^{*} 
& = & 
(1-\thetaCurr )\phiCurr^{*}
+ \thetaCurr f(\vyCurr)
-\frac{\thetaCurr^{2}}{2\zetaNext \probCurr^2}
\norm{\gradfikVec(\vyCurr)}_{\pseudo{\A}}^2 
\nonumber 
\\
&  & 
+
\frac{\thetaCurr(1 - \thetaCurr )\zetaCurr }{\zetaNext}
\left(\frac{\sigmaA}{2} \normA{\vyCurr-\vvCurr}^{2}
+ \frac{1}{\probCurr}
\innerprod{\pseudo{\A}\gradfikVec(\vyCurr)}{\vvCurr - \vyCurr}_{\A}\right)
\enspace.\nonumber 
\end{eqnarray}
\end{lemma}

\begin{proof}\footnote{Note that our proof was heavily influenced by the proof in \cite{Nesterov2003} for estimate sequences.}
First prove by induction on $k$ that $\E[\phi_k(\vx)] \leq (1 - \eta_k) f(\vx) + \eta_k \E[\phi_0(\vx)]$. The base case, $k = 0$, follows trivially from $\etaInit = 1$. Assuming it holds for some $k$, we explicitly writing out the expectation over $\indexCurr$ yields that the following holds for all $\vx \in \Rn$
\[
\E[\phi_{k + 1}(\vx)]
=
\Eof{\sum_{\probCurr} \probCurr \left(
(1 - \thetaCurr) \phiCurr(\vx) + \thetaCurr
\left[
f(\vyCurr) + \probCurr^{-1} \innerprod{\gradfikVec(\vyCurr)}{\vx - \vyCurr}
+ \frac{\sigmaA}{2} \normA{\vx - \vyCurr}^{2}
\right]
\right)
}.
\]
Applying the inductive hypothesis, we get
\[
\E[\phi_{k + 1}(\vx)]
=
(1 - \thetaCurr) \left[(1 - \etaCurr) f(\vx) + \etaCurr \Eof{\phiInit(\vx)}\right]  + 
\thetaCurr
\Eof{
f(\vyCurr) + \innerprod{\grad f(\vyCurr)}{\vx - \vyCurr}
+ \frac{\sigmaA}{2} \normA{\vx - \vyCurr}^{2}} .
\]
By strong convexity and the definition of $\etaNext$, we then have that
\[
\E[\phi_{k + 1}(\vx)]
\leq
(1 - \thetaCurr) (1 - \etaCurr) f(\vx) 
+  \thetaCurr f(\vx)
+ (1 - \thetaCurr ) \etaCurr \Eof{\phiInit(\vx)}
=
(1 - \etaNext) f(\vx) 
+ \etaNext \Eof{\phiInit(\vx)} .
\]
Next , we prove the form of $\phiCurr$ again by induction on $k$. Suppose that $\phiCurr = \phiCurr^* + \frac{\zetaCurr}{2} \normA{\vx - \vv_k}^2$. Now, we prove the form of $\phi_{k + 1}$. By the update rule and the inductive hypothesis, we see that
\[
\forall \vx \in \Rn
\enspace : \enspace
\hessian \phiNext(\vx)
= (1 - \thetaCurr) \hessian \phiCurr(\vx) + \thetaCurr \sigmaA \A
= \left[ (1 - \theta_k) \zeta_k + \thetaCurr \sigma_{\A} \right] \A
\enspace.
\]
Consequently, $\phiNext(\vx) = \phiNext^* + \frac{\zetaNext}{2} \normA{\vx - \vvNext}^2$ for $\zetaNext$ as desired and $\phiNext^*$ and $\vvNext$ yet to be determined.

To compute $\vvNext$, we note that $\grad \phiCurr(\vx) = \zetaCurr \A (\vx - \vvCurr)$ and therefore by the update rule 
\begin{align*}
\grad \phiNext(\vx) 
&= (1 - \thetaCurr) \zetaCurr \A (\vx - \vv_k)
+ \frac{\theta_k}{p_{i_k}} \gradfikVec
+ \theta_k \sigma_{\A} \A (\vx - \vy_k)
\\
&= 
\zeta_{k + 1} \A \vx - (1 - \theta_k) \zeta_k \A \vv_k
+ \frac{\theta_k}{p_{i_k}} \gradfikVec
- \thetaCurr \sigma_{\A} \A \vy_k
\end{align*}
Therefore, we have that $\vv_{k + 1}$ must satisfy
\[
\zeta_{k + 1} \A \vx - \zeta_{k + 1} \A \vv_{k + 1}
=
\zeta_{k + 1} \A \vx - (1 - \theta_k) \zeta_k \A \vv_k
+ \frac{\theta_k}{p_{i_k}} \gradfikVec(\vyCurr)
- \theta \sigma_{\A} \A \vy_k.
\]
So, applying $\pseudo{\A}$ to each side yields the desired formula for $\vvNext$.

Finally, to compute $\phiNext^*$, we note that, by the form of $\phiNext(\vx)$ and the update rule for creating $\phiNext(\vx)$, looking at $\phiNext(\vyCurr)$ yields
\[
\phiNext^* + \frac{\zetaNext}{2} \normA{\vyCurr - \vvNext}^2
= (1 - \thetaCurr) \phiCurr(\vyCurr) + \thetaCurr f(\vyCurr)
=
(1 - \thetaCurr) \left[\phiCurr^* + \frac{\zetaCurr}{2} \normA{\vyCurr - \vvCurr}^2 \right]
+ \thetaCurr f(\vyCurr).
\]
Now, by value of $v_{k + 1}$, we see that
\begin{align*}
\zetaNext^2 \normA{\vvNext - \vyCurr}^2
&= \normA{(1 - \thetaCurr) \zetaCurr \vvCurr + (\thetaCurr \sigmaA - \zetaNext) \vyCurr - \frac{\thetaCurr}{\probCurr} \pseudo{\A} \gradfikVec(\vyCurr)}^2 \\
&= 
\normA{(1 - \thetaCurr) \zetaCurr (\vvCurr - \vyCurr) - \frac{\thetaCurr}{\probCurr} \pseudo{\A} \gradfikVec(\vyCurr)}^2
\\
&= 
(1 - \thetaCurr)^2 \zetaCurr^2 \normA{\vvCurr - \vyCurr}^2
-\frac{2 (1 - \thetaCurr) \thetaCurr \zetaCurr}{\probCurr}
 \innerprod{\pseudo{\A} \gradfikVec(\vyCurr)}{\vvCurr - \vyCurr}_{\A}
 +
\frac{\thetaCurr^2}{\probCurr^2} \normA{\pseudo{\A} \gradfikVec(\vyCurr)}^2 .
\end{align*}
Combining these two formulas and noting that
\[
(1 - \thetaCurr) \frac{\zetaCurr}{2} -
\frac{\zetaNext}{2} \cdot \frac{1}{\zetaNext^2}(1- \thetaCurr)^2 \zetaCurr^2
= 
\frac{(1 - \thetaCurr) \zetaCurr}{2 \zetaNext} \left[\zetaNext - (1 - \thetaCurr)\zetaCurr\right]
= \frac{\thetaCurr (1 - \thetaCurr) \zetaCurr}{\zetaNext} \cdot \frac{\sigmaA}{2}
\]
yields the desired form of $\phiNext^*$ and completes the proof.
\end{proof}
\section{Bounding ACDM Coefficients}
\label{sec:app:acdm_coefficients}

In the following lemma we prove bounds on $\alpha_{k}, \beta_{k}, \gamma_{k}, a_{k}$
and $b_{k}$ required for Theorem \ref{thm:main_result} and Lemma \ref{lem:stable}. 

\begin{lemma}[ACDM Coefficients]
\label{lem:coeff}
In ACDM $\gamma_{k}$ is increasing to $\sqrt{\frac{\tilde{S}_{\alpha}}{2n\sigmaOma}}$ and $\beta_{k}$ is decreasing to $1-\sqrt{\frac{\sigmaOma}{2\tsa n}}$. Furthermore, for all $k \geq 1$ we have 
$
\alpha_{k}\leq 32 \max\left(\frac{1}{k},\sqrt{\frac{\sigmaOma}{2\tsa n}}\right)
$
and for all $k \geq 0$ we have
\begin{eqnarray}
a_{k} & \geq & \sqrt{\frac{\tsa}{2n\sigmaOma}}\left(\left(1+\frac{1}{2}\sqrt{\frac{\sigmaOma}{2\tsa n}}\right)^{k+1}-\left(1-\frac{1}{2}\sqrt{\frac{\sigmaOma}{2\tsa n}}\right)^{k+1}\right),
\label{eq:ak_growth}
\\
b_{k} & \geq & \left(\left(1+\frac{1}{2}\sqrt{\frac{\sigmaOma}{2\tsa n}}\right)^{k+1}+\left(1-\frac{1}{2}\sqrt{\frac{\sigmaOma}{2\tsa n}}\right)^{k+1}\right)
\label{eq:bk_growth}
\enspace.
\end{eqnarray}
\end{lemma}
\begin{proof}
We begin by estimating $\gamma_k$. Using that $\gamma_{k} = a_{k + 1} / b_{k + 1}$ and defining $\gamma_{-1} \defeq a_0 / b_0 = 1 /(4n)$ we know that 
$
\gamma_{k+1}^{2}-\frac{\gamma_{k+1}}{2n}=\left(1-\frac{\gamma_{k+1}\sigmaOma}{\tsa}\right)\gamma_{k}^{2}
$
and therefore
$
\frac{\tsa}{2n\sigmaOma}-\gamma_{k+1}^{2}=\left(1-\frac{\gamma_{k+1}\sigmaOma}{\tsa}\right)\left(\frac{\tsa}{2n\sigmaOma}-\gamma_{k}^{2}\right)
$
for all $k \geq 0$. Consequently, $\gamma_{k}$ is increasing to $\sqrt{\frac{\tilde{S}_{\alpha}}{2n\sigmaOma}}$ and since $\beta_{k}=1-\frac{\gamma_{k}\sigmaOma}{\tsa}$, we have that $\beta_{k}$ is decreasing to $1-\sqrt{\frac{\sigmaOma}{2\tsa n}}$. Furthermore since $2 \tsa n \geq \sigmaOma$ we have that all $\beta_k \in (0, 1)$, $b_k \leq b_{k + 1}$, and $a_k \leq a_{k + 1}$.

Using these insights we can now estimate the growth of coefficients $a_{k}$ and $b_{k}$. As in \cite{Nesterov2012} the growth can be estimated by a recursive
relation between $a_{k}$ and $b_{k}$. For $b_{k}$, our definitions imply
\[
b_{k}^{2}
= \beta_{k}b_{k+1}^{2}
= \left(1-\frac{\sigmaOma}{\tsa}\gamma_{k}\right)b_{k+1}^{2}
= \left(1-\frac{\sigmaOma}{\tsa}\frac{a_{k+1}}{b_{k+1}}\right) b_{k+1}^{2}
\enspace.
\]
Therefore since $b_k \leq b_{k + 1}$ we have $\frac{\sigmaOma}{\tsa} a_{k+1}b_{k+1} \leq b_{k+1}^{2}-b_{k}^{2} \leq 2b_{k+1}(b_{k+1}-b_{k})$ and consequently
\begin{equation}
b_{k+1}\geq b_{k}+\frac{\sigmaOma}{2\tsa}a_{k+1}
\enspace.\label{eq:b_inq}
\end{equation}
To bound $a_{k}$, the definitions imply
$
\frac{a_{k+1}^{2}}{b_{k+1}^{2}}-\frac{a_{k+1}}{2nb_{k+1}}=\gamma_{k}^{2}-\frac{\gamma_{k}}{2n}=\beta_{k}\frac{a_{k}^{2}}{b_{k}^{2}}=\frac{a_{k}^{2}}{b_{k+1}^{2}}
$.
Therefore since $a_k \leq a_{k + 1}$ we have
$
\frac{1}{2n}a_{k+1}b_{k+1}=a_{k+1}^{2}-a_{k}^{2}\leq2a_{k+1}(a_{k+1}-a_{k})
$
and consequently
\begin{equation}
a_{k+1} 
\geq a_{k} + \frac{1}{4n}b_{k+1}
\geq a_{k} + \frac{1}{4n}b_{k}
\enspace.\label{eq:a_inq}
\end{equation}
Using (\ref{eq:a_inq}) and (\ref{eq:b_inq}), it is easy to prove the growth rates \eqref{eq:ak_growth} and \eqref{eq:bk_growth} by induction.

All the remains to prove the upper bound on $\alpha_{k}$. Note that 
$
\frac{1-\alpha_{k}}{\alpha_{k}}=\frac{2n\gamma_{k}-1}{\beta_{k}}
$
and hence $\alpha_{k}\sim\frac{\beta_{k}}{2n\gamma_{k}-1}\sim\frac{1}{2n\gamma_{k}}$.
Therefore, we wish to find a lower bound of $\gamma_{k}$,
which in turn can be found by a upper bound of $\beta_{k}$. 
Recalling that $b_{k + 1} = \frac{b_k}{\sqrt{\beta_k}}$ we see that
$
b_{k}^{2}=\frac{b_{0}^{2}}{\beta_{0}\cdots\beta_{k-1}}\leq\frac{b_{0}^{2}}{\beta_{k-1}^{k}}.
$
Therefore, by our lower bound on $b_k$ and the fact that $b_0 = 2$, we see that when $k\leq 2\sqrt{\frac{2\tsa n}{\sigmaOma}}$, we have 
\begin{align*}
\beta_{k} 
&\leq 2^{2/k}\left(\left(1+\frac{1}{2}\sqrt{\frac{\sigmaOma}{2\tsa n}}\right)^{k}+\left(1-\frac{1}{2}\sqrt{\frac{\sigmaOma}{2\tsa n}}\right)^{k}\right)^{-2/k} 
\leq 2^{2/k}\left(2+\frac{\sigmaOma}{8\tsa n}k(k-1)\right)^{-2/k}
\\
&= \left(1+\frac{\sigmaOma}{16\tsa n}k(k-1)\right)^{-2/k}
\leq \exp\left(-\frac{\sigmaOma}{8\tsa n}(k-1)\right)
\leq 1-\frac{\sigmaOma}{16\tsa n}(k-1).
\end{align*}
Using $\beta_{k}=1-\frac{\gamma_{k}\sigmaOma}{\tsa}$, we
get 
$
\gamma_{k}\geq\frac{1}{16}\min\left(\frac{k-1}{n},\sqrt{\frac{2\tsa}{\sigmaOma n}}\right).
$
Therefore since $\frac{1}{\alpha_k} - 1 = \frac{1 - \alpha_k}{\alpha_k} = \frac{n \gamma_k - 1}{\beta_k} \geq n \gamma_k - 1$ for all $k$ we have that $\alpha_k \leq \frac{1}{n \gamma_k} \leq 16 \max \left\{\frac{1}{k - 1}, \sqrt{\frac{\sigmaOma}{2 \tsa n}}\right\}$ and the result follows immediately.
\end{proof}

\section{Proof of Numerically Stable ACDM}
\label{sec:app:num_stab}

\begin{proof}[Proof of Theorem \ref{thm:main_result}]
Following the structure in \cite{Nesterov2012} we begin by defining $T_{i}(\vx) \defeq \vx - \frac{1}{L_{i}} \gradfiVec(\vx)$ for all $\vx \in \Rn$ and $r_{k}^{2} \defeq \normOma{\vvCurr- \xopt}^{2}$ for all $k \geq 0$. Expanding $\vvCurr$ yields
\begin{eqnarray}
r_{k+1}^{2} 
& = & 
||\verrTwoCurr+\beta_{k}\vvCurr+(1-\beta_{k})\vyCurr-\frac{\gamma_{k}}{\tlik}\vec{f_{i_{k}}}(\vyCurr)-\xopt||_{1-\alpha}^{2}\nonumber \\
& \leq & 
\left(1+\frac{1}{t}\right)||\verrTwoCurr||_{1-\alpha}^{2}+(1+t)||\beta_{k}\vvCurr+(1-\beta_{k})\vyCurr-\frac{\gamma_{k}}{\tlik}\vec{f_{i_{k}}}(\vyCurr)-\xopt||_{1-\alpha}^{2}\nonumber \\
& \leq & 
\left(1+\frac{1}{t}\right)\varepsilon^{2}+(1+t)||\beta_{k}\vvCurr+(1-\beta_{k})\vyCurr-\xopt||_{1-\alpha}^{2}+(1+t)\frac{\gamma_{k}^{2}}{\tlik^{1+\alpha}}f_{i_{k}}(\vyCurr)^{2}\nonumber \\
&  & 
+2\frac{\gamma_{k}}{\tlik^{\alpha}}(1+t)\left\langle \vec{f_{i_{k}}}(\vyCurr),\xopt-\beta_{k}\vvCurr-(1-\beta_{k})\vyCurr\right\rangle \nonumber \\
& \leq & 
\left(1+\frac{1}{t}\right)\varepsilon^{2}+(1+t)||\beta_{k}\vvCurr+(1-\beta_{k})\vyCurr-\xopt||_{1-\alpha}^{2}+2\frac{\gamma_{k}^{2}}{\tlik^{\alpha}}(1+t)\left(f(\vyCurr)-f(T_{i_{k}}(\vyCurr))\right)\nonumber \\
&  & 
+2\frac{\gamma_{k}}{\tlik^{\alpha}}(1+t)\left\langle \vec{f_{i_{k}}}(\vyCurr),\xopt-\vyCurr+\frac{\beta_{k}(1-\alpha_{k})}{\alpha_{k}}\left(\vxCurr-\vyCurr\right)\right\rangle ,\label{eq:r_est_1}
\end{eqnarray}
where $t$ is yet to be determined. 

Applying the convexity parameter and
Lipschitz condition, we can bound the error induced by $\verrOneCurr$:
\begin{align}
f(T_{i_{k}}(\vyCurr)) 
&= 
f(\vxNext-\verrOneCurr)  
\tag{Assumption of Step 3d}
\\
&\geq 
f(\vxNext)-\left\langle \verrOneCurr,\grad f(\vxNext)\right\rangle 
\tag{Convexity of $f$}
\\
&= 
f(\vxNext) -
\innerprod{\verrOneCurr}{\grad f(\vxNext)-\grad f(\xopt)}
\tag{Optimality of $\xopt$ implies $\grad f(\xopt) = \vzero$}
\\
&\geq 
f(\vxNext) - 
\normOma{\verrOneCurr}
\normOmaDual{\grad f(\vxNext)-\grad f(\xopt)}
\tag{Cauchy Schwarz}
\\
&\geq 
f(\vxNext)-\varepsilon\tsa||\vxNext-\xopt||_{1-\alpha}
\tag{Lemma 2 of \cite{Nesterov2012}}\\
&\geq 
f(\vxNext)-\frac{\varepsilon\tsa}{\sigmaOma}\left(f(\vxNext)-f^{*}\right).\label{eq:err_induced_by_1}
\end{align}
Taking the expectation of both sides of (\ref{eq:r_est_1}) in $i_{k}$
and using (\ref{eq:err_induced_by_1}) and $\tlik^{\alpha}\geq\frac{\tilde{S}_{\alpha}}{2n}$,
we have
\begin{eqnarray*}
\EikOf{(r_{k+1}^{2})}
& \leq & 
\left(1+\frac{1}{t}\right)\varepsilon^{2}+(1+t)\beta_{k}r_{k}^{2}+(1+t)(1-\beta_{k})||\vyCurr-\xopt||_{1-\alpha}^{2}\\
 &  & +4\frac{\gamma_{k}^{2}n}{\tsa}(1+t)\left(f(\vyCurr)- \EikOf{f(\vxNext)} +\frac{\varepsilon\tsa}{\sigmaOma}\left(\EikOf{f(\vxNext)}-f^{*}\right)\right)\\
 &  & +2(1+t)\frac{\gamma_{k}}{\tsa}\left\langle \grad f(\vyCurr),\xopt-\vyCurr+\frac{\beta_{k}(1-\alpha_{k})}{\alpha_{k}}\left(\vxCurr-\vyCurr\right)\right\rangle \\
 & \leq & \left(1+\frac{1}{t}\right)\varepsilon^{2}+(1+t)\beta_{k}r_{k}^{2}+(1+t)(1-\beta_{k})||\vyCurr-\xopt||_{1-\alpha}^{2}\\
 &  & +4\frac{\gamma_{k}^{2}n}{\tsa}(1+t)
 \left(f(\vyCurr)- \EikOf{f(\vxNext)} +\frac{\varepsilon\tsa}{\sigmaOma}\left(\EikOf{f(\vxNext)} - f^{*}\right)\right)\\
 &  & +2(1+t)\frac{\gamma_{k}}{\tsa}\left(f^{*}-f(\vyCurr)-\frac{1}{2}\sigmaOma||\vyCurr-\xopt||_{1-\alpha}^{2}+\frac{\beta_{k}(1-\alpha_{k})}{\alpha_{k}}\left(f(\vxCurr)-f(\vyCurr)\right)\right).
\end{eqnarray*}
Now since we defined coefficients so the following hold
\[
1-\beta_{k} = \frac{\gamma_{k}}{\tilde{S}_{\alpha}}\sigmaOma
\enspace \text{ and } \enspace
\gamma_{k}^{2}-\frac{\gamma_{k}}{2n}=\frac{\gamma_{k}\beta_{k}(1-\alpha_{k})}{2n\alpha_{k}},
\]
we see that the terms for $\normOma{\vyCurr-\xopt}^{2}$ and $f(\vyCurr)$ cancel and we obtain 
\begin{eqnarray*}
\EikOf{r_{k+1}^{2}}
& \leq & \left(1+\frac{1}{t}\right)\varepsilon^{2}+(1+t)\beta_{k}r_{k}^{2}-4(1+t)\frac{\gamma_{k}^{2}n}{\tsa}
\left(\EcurrOf{f(\vxNext)}-\frac{\varepsilon\tsa}{\sigmaOma}\left(\EikOf{f(\vxNext)}-f^{*}\right)\right)\\
 &  & +2(1+t)\frac{\gamma_{k}}{\tsa}\left(f^{*}+\frac{\beta_{k}(1-\alpha_{k})}{\alpha_{k}}f(\vxCurr)\right)
 \enspace.
\end{eqnarray*}
Multiplying both sides by $\frac{\tsa}{n}b_{k+1}^{2}$ and recalling the following 
\[
b_{k+1}^{2}=\frac{1}{\beta_{k}}b_{k}^{2},\quad a_{k+1}^{2}=\gamma_{k}^{2}b_{k+1}^{2},\quad\gamma_{k}^{2}-\frac{\gamma_{k}}{2n}=\beta_{k}\frac{a_{k}^{2}}{b_{k}^{2}}=\frac{\gamma_{k}\beta_{k}(1-\alpha_{k})}{2n\alpha_{k}},
\]
we get
\begin{eqnarray*}
\frac{\tsa}{n}b_{k+1}^{2}E_{i_{k}}(r_{k+1}^{2}) & \leq & \left(1+\frac{1}{t}\right)\frac{\tsa}{n}b_{k+1}^{2}\varepsilon^{2}+(1+t)\frac{\tsa}{n}b_{k}^{2}r_{k}^{2}\\
 &  & +(1+t)\left(-4a_{k+1}^{2}\left(1-\frac{\varepsilon\tsa}{\sigmaOma}\right)\left(E_{i_{k}}f(\vxNext)-f^{*}\right)+4a_{k}^{2}\left(f(\vxCurr)-f^{*}\right)\right).
\end{eqnarray*}
Dropping the expectation in each variable we have that in expectation up to iteration $k$
\begin{eqnarray*}
 &  & \frac{\tsa}{n}b_{k+1}^{2}(r_{k+1}^{2})+(1+t)\left(1-\frac{\varepsilon\tsa}{\sigmaOma}\right)4a_{k+1}^{2}\left(f(\vxNext)-f^{*}\right)\\
 & \leq & \left(1+\frac{1}{t}\right)\frac{\tsa}{n}b_{k+1}^{2}\varepsilon^{2}+(1+t)\left(\frac{\tsa}{n}b_{k}^{2}r_{k}^{2}+4a_{k}^{2}\left(f(\vxCurr)-f^{*}\right)\right).
\end{eqnarray*}
Letting
$
t\defeq\frac{2\varepsilon\tilde{S}_{\alpha}}{\sigmaOma}.
$
and writing $\tilde{r}_{k}^{2}=\frac{\tilde{S}_{\alpha}}{n} E_{i_{k-1}}[r_{k}^{2}]$,
$\phi_{k} = \EikOf{f(\vxNext)} - \optValue$, we have
\begin{eqnarray}
b_{k+1}^{2}\tilde{r}_{k+1}^{2}+4a_{k+1}^{2}\phi_{k+1} & \leq & \frac{\tsa}{n}\left(1+\frac{1}{t}\right)b_{k+1}^{2}\varepsilon^{2}+(1+t)\left(b_{k}^{2}\tilde{r}_{k}^{2}+4a_{k}^{2}\phi_{k}\right)\nonumber \\
 & \leq & \frac{\tsa}{n}\varepsilon^{2}\left(1+\frac{1}{t}\right)\sum_{j=1}^{k+1}(1+t)^{k+1-j}b_{j}^{2}+4(1+t)^{k+1}\left(\tilde{r}_{0}^{2}+\frac{1}{2n\tsa}\phi_{0}\right).\label{eq:r_est_2}
\end{eqnarray}
Now, we claim the following inequalities:
\[
\begin{cases}
(1+t)^{k+1-j}b_{j}^{2}\text{ is increasing} & (1)\\
\sqrt{\frac{\tsa}{2n\sigmaOma}}\geq\gamma_{k}\geq\frac{1}{2n} & (2)\\
a_{k}\geq\frac{1}{2}\sqrt{\frac{\tsa}{2n\sigmaOma}}\left(1+\frac{1}{2}\sqrt{\frac{\sigmaOma}{2\tsa n}}\right)^{k+1} & (3)
\end{cases}.
\]
Using the claims above and \eqref{eq:r_est_2}, we get
\begin{eqnarray*}
 &  & E_{i_{k}}\left(2\sigmaOma||\vvNext-\xopt||_{1-\alpha}^{2}+4(f(\vxNext)-f^{*})\right)\\
 & = & \frac{2n\sigmaOma}{\tsa}\tilde{r}_{k+1}^{2}+4\phi_{k+1}\\
 & \leq & \frac{b_{k+1}^{2}}{a_{k+1}^{2}}\tilde{r}_{k+1}^{2}+4\phi_{k+1}\\
 & \leq & \frac{\tsa}{n}\varepsilon^{2}\left(1+\frac{1}{t}\right)(k+1)\frac{b_{k+1}^{2}}{a_{k+1}^{2}}+\frac{4(1+t)^{k+1}}{a_{k+1}^{2}}\left(\tilde{r}_{0}^{2}+\frac{1}{2n\tsa}\phi_{0}\right)\\
 & \leq & 
 6\tsa\varepsilon^{2}(k+1)+32\frac{n\sigmaOma}{\tsa}\left(1+\frac{2\varepsilon\tsa}{\sigmaOma}
 \right)^{k+1}\left(1+\frac{1}{2}\sqrt{\frac{\sigmaOma}{2\tsa n}}\right)^{-2(k+1)}\left(\tilde{r}_{0}^{2}+\frac{1}{2n\tsa}\phi_{0}\right).
\end{eqnarray*}
By the assumption that $\frac{2\varepsilon\tsa}{\sigmaOma}<\frac{1}{2}\sqrt{\frac{\sigmaOma}{2\tsa n}}$
inequality (\ref{eq:ACD_bound}) follows from the following
\begin{eqnarray*}
 &  & E_{i_{k}}\left(2\sigmaOma||\vvNext-\xopt||_{1-\alpha}^{2}+4(f(\vxNext)-f^{*})\right)\\
 & \leq & 12k\tsa\varepsilon^{2}+32\sigmaOma\left(1+\frac{1}{2}\sqrt{\frac{\sigmaOma}{2\tsa n}}\right)^{-k}\left(\normOma{\vxInit -  \xopt}^{2}+\frac{1}{2\tsa^{2}}\left(f(x_{0})-f^{*}\right)\right)\\
 & \leq & 24kS_{\alpha}\varepsilon^{2}+32\sigmaOma\left(1-\frac{1}{5}\sqrt{\frac{\sigmaOma}{S_{\alpha}n}}\right)^{k}\left(\normOma{\vxInit -  \xopt}^{2}+\frac{1}{S_{\alpha}^{2}}\left(f(x_{0})-f^{*}\right)\right).
\end{eqnarray*}
Now, we prove the claims. 

Claim (1): Since $a_{k+1} = \gamma_{k}b_{k+1}\geq\frac{1}{2n}b_{k+1}$
and (\ref{eq:b_inq}), we have
$
b_{k+1}\geq(1+\frac{\sigmaOma}{4\tilde{S}_{\alpha}n})b_{k}
$.
By the assumption $t=\frac{2\varepsilon\tilde{S}_{\alpha}}{\sigmaOma}<\frac{\sigmaOma}{4\tilde{S}_{\alpha}n}$,
we see that $(1+t)^{k+1-j}b_{j}^{2}$ is increasing. 

Claim (2): Follows from Lemma \ref{lem:coeff}. 

Claim (3): Using Lemma \ref{lem:coeff} and $k\geq\sqrt{\frac{\sigmaOma}{2\tsa n}}$,
we have
\begin{eqnarray*}
a_{k} 
& \geq & 
\sqrt{\frac{\tsa}{2n\sigmaOma}}
\left(\left(1+\frac{1}{2}\sqrt{\frac{\sigmaOma}{2\tsa n}}\right)^{k+1}-\left(1-\frac{1}{2}\sqrt{\frac{\sigmaOma}{2\tsa n}}\right)^{k+1}\right)\\
& \geq & 
\sqrt{\frac{\tsa}{2n\sigmaOma}}\left(\left(1+\frac{1}{2}\sqrt{\frac{\sigmaOma}{2\tsa n}}\right)^{k+1}-1\right)\\
& \geq & \frac{1}{2}\sqrt{\frac{\tsa}{2n\sigmaOma}}\left(1+\frac{1}{2}\sqrt{\frac{\sigmaOma}{2\tsa n}}\right)^{k+1}.
\end{eqnarray*}

To bound the norm of the gradient, we show that if $\normOma{\gradient f(\vyCurr)}^{2}$
is large for many steps then $f(\vxCurr)$ decreases substantially. For notational simplicity we omit the expectation symbol and using \eqref{eq:err_induced_by_1},
we have
\begin{eqnarray*}
f(\vxNext) & \leq & f(T_{i_{k}}(\vyCurr))+\frac{\varepsilon\tsa}{\sigmaOma}\left(f(\vxNext)-f^{*}\right)\\
 & \leq & f(\vyCurr)-\frac{1}{\tsa}\left(||\grad f(\vyCurr)||_{1-\alpha}^{*}\right)^{2}+\frac{\varepsilon\tsa}{\sigmaOma}\left(f(\vxNext)-f^{*}\right).
\end{eqnarray*}
To bound $f(\vyCurr)$, we consider the lower envelops at $\vyCurr$ and
obtain 
\begin{eqnarray*}
f(\vyCurr) & \leq & f(\vxCurr)-\left\langle \gradient f(\vyCurr),\alpha_{k}(\vxCurr-\vvCurr)\right\rangle \\
 & \leq & f(\vxCurr)+\alpha_{k}||\gradient f(\vyCurr)||_{1-\alpha}^{*}||\vxCurr-\vvCurr||_{1-\alpha}\\
 & \leq & f(\vxCurr)+\alpha_{k}||\gradient f(\vyCurr)||_{1-\alpha}^{*}\left(||\vxCurr-\xopt||_{1-\alpha}+||\vvCurr-\xopt||_{1-\alpha}\right)\\
 & \leq & f(\vxCurr)+\alpha_{k}||\gradient f(\vyCurr)||_{1-\alpha}^{*}\left(\frac{1}{\sqrt{\sigmaOma}}\sqrt{f(\vxCurr)-f^{*}}+||\vvCurr-\xopt||_{1-\alpha}\right).
\end{eqnarray*}
Combining with Lemma \ref{lem:coeff}, we have
\[
f(\vxNext) 
\leq f(\vxCurr)
- \frac{1}{\tsa}\left(\normOmaDual{\grad f(\vyCurr)}\right)^{2}
+ \frac{\varepsilon\tsa}{\sigmaOma}\delta_{k+1}
+ 3 \sqrt{\frac{\sigmaOma}{n\tsa}} \normOmaDual{\grad f(\vyCurr)}\sqrt{\frac{\delta_{k}}{\sigmaOma}}
\enspace.
\]
Summing up from $k$ to $2k$ then yields
\[
f(\vx_{2k}) 
\leq f(\vxCurr)
- \frac{1}{\tsa}\sum_{j=k}^{2k-1}\left(||\grad f(\vyCurr)||_{1-\alpha}^{*}\right)^{2}
+ \frac{\varepsilon\tsa}{\sigmaOma}\sum_{j=k+1}^{2k}\delta_{j}
+ 32\sqrt{\frac{1}{n\tsa}}\sum_{j=k}^{2k-1} \normOmaDual{\grad f(\vy_{j})}\sqrt{\delta_{j}}
\enspace.
\]
Since $f(\vxCurr)-f(\vx_{2k})\leq f(\vxCurr)-f^{*}\leq\delta_{k}$, we have
\begin{eqnarray*}
\frac{1}{\tsa}\sum_{j=k}^{2k-1}\left(||\grad f(\vy_j)||_{1-\alpha}^{*}\right)^{2} 
& \leq & \delta_{k}+\frac{\varepsilon\tsa}{\sigmaOma}\sum_{j=k+1}^{2k}\delta_{j}+32\sqrt{\frac{1}{n\tsa}}
\sum_{j=k}^{2k-1} \normOmaDual{\grad f(\vy_{j})} \sqrt{\delta_{j}}
\\
 & \leq & 
 \delta_{k}\left(1+\frac{\varepsilon k\tsa}{\sigmaOma}\right)+32\sqrt{\frac{1}{n\tsa}}\sum_{j=k}^{2k-1} \normOmaDual{\grad f(\vy_{j})} \sqrt{\delta_{j}}
 \\
 & \leq & 
 \delta_{k}\left(1+\frac{\varepsilon k\tsa}{\sigmaOma}\right)+32\sqrt{\frac{k\delta_{k}}{n}}\sqrt{\frac{1}{\tilde{S}_{\alpha}}
 \sum_{j=k}^{2k-1}\left( \normOmaDual{\grad f(\vyCurr)}\right)^{2}}.
\end{eqnarray*}
Solving this quadratic equation about $\sqrt{\frac{1}{\tsa}\sum_{j=k}^{2k-1}\left(\normOmaDual{\grad f(\vyCurr)}\right)^{2}}$,
we get
\begin{eqnarray*}
\sqrt{\frac{1}{\tsa}\sum_{j=k}^{2k-1}\left(||\grad f(\vyCurr)||_{1-\alpha}^{*}\right)^{2}} 
& \leq & 
\frac{1}{2} \left[32\sqrt{\frac{k\delta_{k}}{n}}+\sqrt{\left(32\sqrt{\frac{k\delta_{k}}{n}}\right)^{2}+4\delta_{k}\left(1+\frac{\varepsilon k\tsa}{\sigmaOma}\right)}\right].
\end{eqnarray*}
Since $1+\frac{\varepsilon k\tsa}{\sigmaOma}\leq2$ and $k \geq n$ by assumptions we have
\[
\sqrt{\frac{1}{\tsa}\sum_{j=k}^{2k-1}\left(
\normOmaDual{\grad f(\vyCurr)}\right)^{2}} 
\leq 16\sqrt{\frac{k\delta_{k}}{n}}
+
\frac{1}{2} \sqrt{(32^2 + 8)\frac{k\delta_{k}}{n}}
\leq 33\sqrt{\frac{k\delta_{k}}{n}}.
\]
Therefore 
$
\frac{1}{\tsa}\sum_{j=k}^{2k-1}\left(||\grad f(\vyCurr)||_{1-\alpha}^{*}\right)^{2} \leq 2000\frac{k\delta_{k}}{n}.
$
as desired.
\end{proof}
\begin{lemma}[Coefficient Stability in ACDM]
\label{lem:coeff_stable} Step 3a of ACDM can be computed by the following formulas
\begin{eqnarray}
\gamma_{k+1} & = & 
\min \left\{
\frac{1}{2} \left[
\left(\frac{1}{2n}-\frac{\gamma_{k}^{2}\sigmaOma}{\tsa}\right)+\sqrt{\left(\frac{1}{2n}-\frac{\gamma_{k}^{2}\sigmaOma}{\tsa}\right)^{2}+4\gamma_{k}^{2}}\right]
\enspace
,
\enspace
\sqrt{\frac{\tilde{S}_{\alpha}}{2n\sigmaOma}}
\right\}
,\label{eq:gamma_formula}
\\
\beta_{k+1} & = & 1-\frac{\gamma_{k+1}\sigmaOma}{\tsa},\nonumber \\
\alpha_{k+1} & = & \frac{\beta_{k+1}}{\beta_{k+1}+2n\gamma_{k+1}-1}
= \frac{\frac{2n}{\gamma_k} - \frac{\sigmaOma}{2n S_\alpha}}{1 - \frac{\sigmaOma}{2n S_\alpha}}
.\nonumber 
\end{eqnarray}
and can be implemented using $O(\log n)$ bits precision to obtain
any $poly(\frac{1}{n})$ additive accuracy.\end{lemma}
\begin{proof}
Since $\gamma_k = \frac{a_{k + 1}}{b_{k + 1}}$ we know that
$
\gamma_{k+1}^{2}-\frac{\gamma_{k+1}}{2n}=\left(1-\frac{\gamma_{k+1}\sigmaOma}{\tsa}\right)\gamma_{k}^{2},
$
and by re-arranging terms we get that
$
\gamma_{k+1}^{2}-\left(\frac{1}{2n}-\frac{\gamma_{k}^{2}\sigmaOma}{\tsa}\right)\gamma_{k+1}-\gamma_{k}^{2}=0.
$
Applying the quadratic formula and Lemma \ref{lem:coeff} then yields \eqref{eq:gamma_formula} and the other formulas follow by direct computation.

Now, expanding (\ref{eq:gamma_formula}) and using $2\tsa n \geq \sigmaOma$ implies that $\gamma_{k} \geq \frac{1}{2n}$ and therefore the denominator of $\alpha_{k+1}$ is larger than $\beta_{k+1}$. Consequently the equations of $\beta$ and $\alpha$ depends continuously on $\gamma$. Thus, it is suffices to show that $O(\log n)$ bits precision
suffice to compute $\gamma_k$. To prove this, we define $f : \Rn \rightarrow \R$ denote the quadratic formula so that with full precision $\gamma_{k+1} = f(\gamma_{k})$. Direct calculation yields
\[
f'(\gamma) 
= 
-\frac{\gamma\sigmaOma}{\tsa}
+ \frac{\left(\frac{\gamma^{2}\sigmaOma}{\tsa} - \frac{1}{2n}\right)
\frac{\gamma\sigmaOma}{\tsa}+2\gamma}{\sqrt{\left(\frac{1}{2n}-\frac{\gamma^{2}\sigmaOma}{\tsa}\right)^{2}+4\gamma^{2}}}
\]
Since $0 \leq \gamma \leq\sqrt{\frac{\tilde{S}_{\alpha}}{2n\sigmaOma}}$ we have that $0 \leq f'(\gamma) \leq 1 - \frac{\gamma \sigmaOma}{\tsa} < 1$ and therefore, $f$ is a contraction mapping. Consequently, if the calculation of $\gamma$ has up to $\epsilon$ additive error, $k$ steps of calculation can accumulate up to $k \epsilon / (1-\max f'(\gamma))$ additive error. Hence, $O(\log n)$ bits of precision suffice.
\end{proof}

\end{document}